\documentclass[useAMS,usenatbib]{mn2e}

\usepackage{graphicx}	% Including figure files
\usepackage{amsmath}	% Advanced maths commands
\usepackage{amssymb}	% Extra maths symbols
\usepackage{upgreek}
\newcommand{\MS}{\ifmmode{\,}\else\thinspace\fi{\rm M}\ifmmode_{\odot}\else$_{\odot}$\fi}
\newcommand{\LS}{\ifmmode{\,}\else\thinspace\fi{\rm L}\ifmmode_{\odot}\else$_{\odot}$\fi}
\newcommand{\RS}{\ifmmode{\,}\else\thinspace\fi{\rm R}\ifmmode_{\odot}\else$_{\odot}$\fi}
\newcommand{\Ke}{\ifmmode{\,}\else\thinspace\fi{\rm K}}

\newcommand{\teff}{\ifmmode T_{\rm eff}\else$T_{\rm eff}$\fi}
\title[Survey of type-II Cepheid models]{Survey of non-linear hydrodynamic models of type-II Cepheids}
\author[R. Smolec]{
R. Smolec\thanks{E-mail: smolec@camk.edu.pl}
\\
Nicolaus Copernicus Astronomical Center, Polish Academy of Sciences, ul. Bartycka 18, 00-716 Warszawa, Poland
}

\begin{document}

\date{Accepted . Received ; in original form }

\pagerange{\pageref{firstpage}--\pageref{lastpage}} \pubyear{2015}

\maketitle

\label{firstpage}

\begin{abstract}
We present a grid on non-linear convective type-II Cepheid models. The dense model grids are computed for $0.6\MS$ and a range of metallicities (${\rm [Fe/H]}\!=\!-2.0,\,-1.5,\,-1.0$), and for $0.8\MS$ (${\rm [Fe/H]}\!=\!-1.5$). Two sets of convective parameters are considered. The models cover the full temperature extent of the classical instability strip, but are limited in luminosity; for the most luminous models violent pulsation leads to the decoupling of the outermost model shell. Hence, our survey reaches only the shortest period RV~Tau domain.

In the Hertzsprung-Russel diagram we detect two domains in which period doubled pulsation is possible. The first extends through the BL~Her domain and low luminosity W~Vir domain (pulsation periods $\sim\!2-6.5$\thinspace d). The second domain extends at higher luminosities (W~Vir domain; periods $>9.5$\thinspace d). Some models within these domains display period-4 pulsation. We also detect very narrow domains ($\sim\!10$\thinspace K wide) in which modulation of pulsation is possible. Another interesting phenomenon we detect is double-mode pulsation in the fundamental mode and in the fourth radial overtone. Fourth overtone is a surface mode, trapped in the outer model layers. Single-mode pulsation in the fourth overtone is also possible on the hot side of the classical instability strip. The origin of the above phenomena is discussed. In particular, the role of resonances in driving different pulsation dynamics as well as in shaping the morphology of the radius variation curves is analysed.
\end{abstract}

\section{Introduction}

Type-II Cepheids are low mass, population II stars, pulsating in the radial fundamental mode. Depending on the pulsation period they are divided into three classes, believed to correspond to different evolutionary stages \citep[although this is based on rather old evolutionary calculations; see e.g.][]{gingold}. BL~Her stars, with periods between 1 and 4\thinspace d, are crossing the instability strip (IS) evolving from the blue part of the Zero-Age Horizontal Branch (ZAHB) towards the Asymptotic Giant Branch (AGB). W~Vir stars, with periods between 4 and 20\thinspace d, loop back into IS as a result of instabilities in two-shell burning along the AGB. Finally, RV~Tau stars, with periods above 20\thinspace d, cross the IS as they evolve away from the AGB. Characteristic feature and discriminant of the RV~Tau class is the phenomenon of period doubling. The light curve is characterized by alternating deep and shallow brightness minima/maxima. The effect seems to appear at periods above 20\thinspace d. The division into three classes is out of necessity arbitrary. Recently it was found that the prototype of the W~Vir class (with $P\approx 17.3$\thinspace d) displays low amplitude alternations \citep{th07} just as is characteristic for RV~Tau class. Type-II Cepheids follow a distinct period-luminosity relation and may be used as distance indicators \citep[e.g.][]{csbook}.

Pulsation dynamics of type-II Cepheids is very interesting. The longer the period, the more irregularities in the light curves appear. Irregular pulsations are observed in W~Vir and RV~Tau classes. In the latter class, irregular pulsations are observed on top of the pronounced period doubling alternations \citep[e.g.][]{t2cep_lmc,blher_ogle}. These phenomena are now better understood thanks to non-linear modelling of pulsation. The largest survey of type-II Cepheid models published so far, was conducted by \cite{kb88} who used purely radiative, lagrangian pulsation code. They reproduced the period doubling effect, which was later understood as arising due to half-integer resonance between the pulsation modes \citep{mb90}. The models of Kov\'acs \& Buchler show so-called period doubling route to chaos; through successive period doubling bifurcations the deterministic chaos is reached. Other route to chaos revealed in their models is due to tangent bifurcation. The period doubling domain starts at pulsation periods around 10\thinspace d in their models, much too short as compared with observations ($\sim$20\thinspace d).

Period-4 pulsations were not clearly revealed in observations yet \citep[see however][our analysis of OGLE-III data for the star they reported does not reveal period-4 behaviour]{pollard}. Based on hydrodynamic modelling however, it seems reasonable that irregularities observed in type-II Cepheids are due to deterministic chaos, which may be reached through a period doubling cascade or through other route. In fact chaos was detected in type-II Cepheids of RV~Tau type \citep[R~Scuti and AC~Her;][]{bks96,kbsm98} and in several semi-regular variables \citep{bkc04}, and in Mira-type variable \citep{ks03}, which are even more luminous.

The strong support for the conclusions arising from relatively  simple, 1D non-linear models came out just recently, with the discovery of period doubling in BL~Her-type star with 2.4\thinspace d pulsation period \citep{blher_ogle,blherPD}. The existence of period doubling effect in BL~Her class was predicted 20 years earlier, based on hydrodynamic calculations of \cite{bm92} \citep[see also][]{ff85}, who found the effect in a series of radiative models with periods between 2 and 2.6\thinspace d. They traced the origin of the phenomenon to the 3:2 resonance between the fundamental mode and the first overtone, which was later confirmed with hydrodynamic calculations including turbulent convection aimed to model the only period doubled BL~Her star currently known \citep{blherPD}.

The systematic survey of type-II Cepheid models with updated microphysics data, convective pulsation codes and realistic model parameters was not conducted since the work of \cite{kb88}, although some limited model surveys were computed, e.g. of BL~Her-type models by \cite{dicriscienzo}. In two recent  papers \cite{blher_mod,blher_chaos} also focused on BL~Her-type models and showed that the models with decreased eddy viscous dissipation display a wealth of dynamical behaviours, from periodic and quasi-periodic modulation of pulsation, akin to the Blazhko effect observed in RR~Lyr stars \citep{blher_mod}, to deterministic chaos with all its features known from textbook chaotic systems, including period doubling and intermittent routes to chaos, stable period-$k$ (even and odd) domains within chaotic bands, type-I and type-III intermittency or crises bifurcation [\cite{blher_chaos}; narrow windows of chaotic oscillations in BL Her models were also reported by \cite{bm92}]. These models, although interesting from dynamical point of view, are characterized by excessive amplitudes (because of the decreased viscous dissipation), as compared to observations, and cannot be regarded a realistic models for the bulk of BL~Her stars or type-II Cepheids. Whether the above phenomena can persist in models with more realistic parameters was not studied so far. It is the goal of the present investigation to analyse the realistic, convective pulsation models of type-II Cepheids with up-to-date microphysics data.

\section{Models}
%%%%%%%%%%%%%%%%%%%%%%%%%%%%%%%%%%%%

All models were computed with the non-linear convective pulsation codes of \cite{sm08a}. These are one-dimensional and lagrangian codes that implement the \cite{kuhfuss} model of turbulent convection. Radiation is treated in the diffusion approximation, which is numerically not expensive. Unfortunately, it is not a good approximation in the outermost layers of the  most luminous type-II Cepheid models. Consequently, we do not expect to get good-looking and smooth light curves, which nicely fit the observed ones. On the other hand, the pulsation dynamics, a subject of this paper, captured through the analysis of radius variation, should be reasonably modelled.

We have computed a grid of models covering the full temperature extent of the instability strip assuming either $M=0.6\MS$ or $0.8\MS$. Model luminosities start at $50\LS$ (RR~Lyr domain) and are increased by $25\LS$ up to $600\LS$ ($0.6\MS$) or $975\LS$ ($0.8\MS$) (see Section~\ref{ssec:inst}). Along each model sequence with constant $L$, models are computed with $25\Ke$ step in effective temperature. This relatively dense grid is designed to detect phenomena that may be restricted to  a narrow parameter ranges. All models adopt OPAL opacity tables \citep{opal} supplemented at low temperatures with \cite{ferguson} opacity data. \cite{asplund} scaled solar mixture is adopted. Three chemical compositions are considered for the $0.6\MS$ models (all with $X=0.75$): $Z=0.0014$ (${\rm [Fe/H]}\approx -1.0$), $Z=0.0004$ (${\rm [Fe/H]}\approx -1.5$) and $Z=0.00014$ (${\rm [Fe/H]}\approx -2.0$); the intermediate chemical composition is adopted for more massive models.

Our models are essentially the same as those in \cite{blherPD}. The model is divided into 150 mass shells extending down to $2\times10^6$\thinspace K. The masses of 110 shells below the anchor zone, which is located at $11\,000$\thinspace K, increase geometrically inward, while shells above the anchor zone have equal mass, except the outermost two shells which are given a factor 2 extra weight (see Section~\ref{ssec:inst}). Our basic set of convective parameters, set A in the following, is the same as in \cite{blherPD} (see tab.~2, set P1, therein) and was used for successful  modelling of the only BL~Her star with period doubling. It assures reasonable pulsation amplitudes and light curve shapes, at least in the least luminous models. We also explore the effect of stronger eddy viscous dissipation, i.e. in set B we set the eddy viscosity parameter to $\alpha_{\rm m}=0.5$ instead of  $\alpha_{\rm m}=0.25$.

Each of the models in our dense model grid was subject to linear non-adiabatic stability analysis. Linear periods of the first five radial modes (fundamental and four consecutive radial overtones) were computed together with the corresponding linear growth rates. Static structure was then used as a starting point for the non-linear model integration, during which the lagrangian model grid was preserved. Scaled velocity eigenvector of the fundamental mode served as an initial perturbation. Model integration was carried for $2\,000$ pulsation cycles (and up to $6\,000$ cycles for the least luminous models close to the boundaries of the IS), each cycle covered with roughly $1\,200$ time steps. As growth rates for type-II Cepheid models are large, typically after few tens up to a few hundred of pulsation cycles the model settled on the finite amplitude pulsation. The longer integration is necessary to check the stability of pulsation and to provide sufficiently long time-series for subsequent analysis.

Except close to the edges of the instability strip, our models are strongly non-linear. Pulsation amplitudes are large and light curve shapes are strongly non-sinusoidal. Pronounced period doubling alternations are present over significant parts of the HR diagram. Significant period changes are also detected in the models. These effects are discussed in detail in Section~\ref{sec:results}.

\subsection{Dynamic instability in the most luminous models}\label{ssec:inst}
%%%%%%%%%%%%%%%%%%%%%%%%%%%%%%%%%%%%%%%%%%%%%%%%%%%%%%%%%%%%%%%%%%%%%%%%%%%%%

\begin{figure}
\centering
\resizebox{\hsize}{!}{\includegraphics{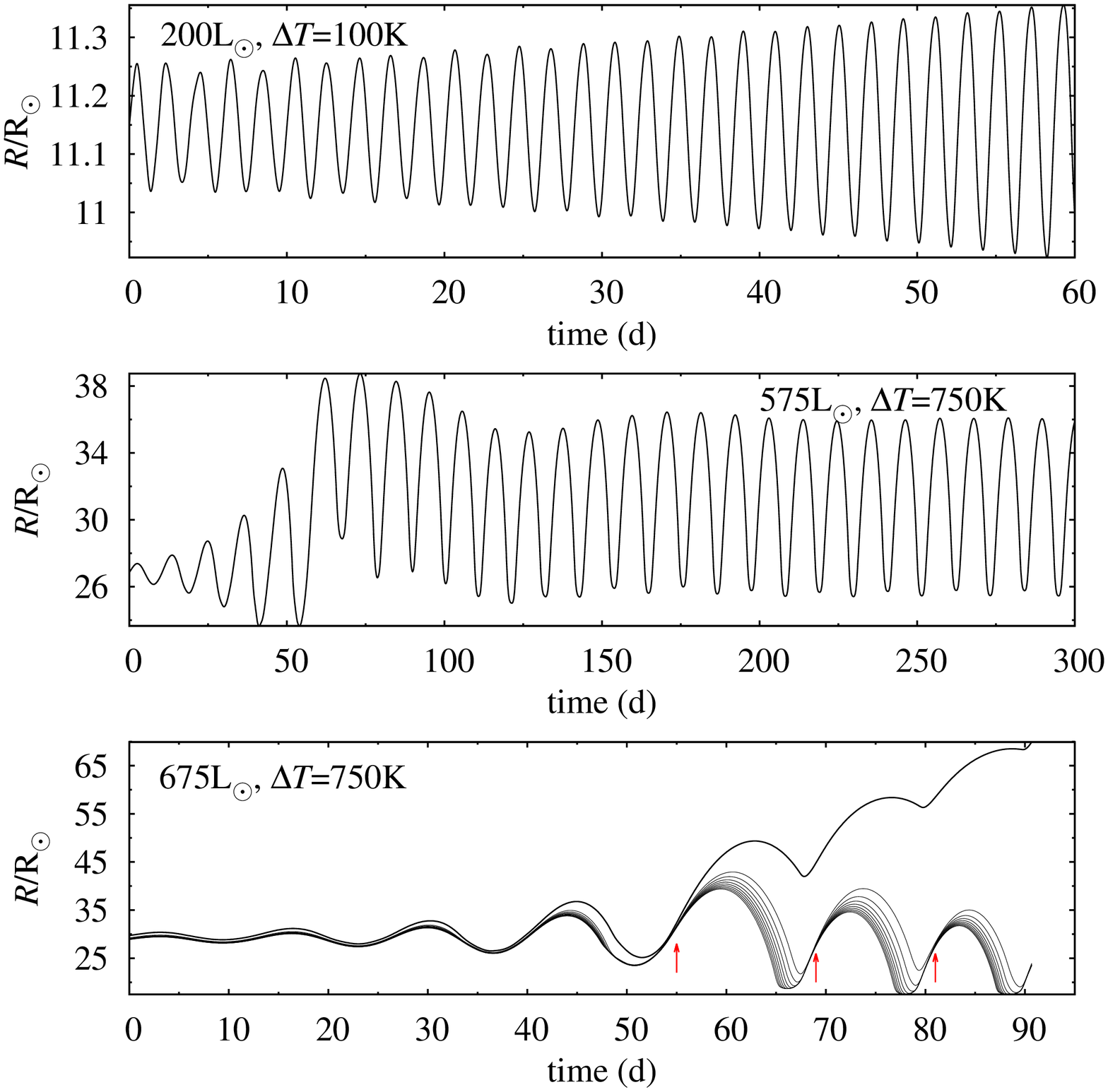}}
\caption{Variation of the radius of the outermost model shell at the initial phase of model integration. In each panel the model's luminosity and the distance from the blue edge of the IS, $\Delta T$, are given. In the bottom panel radii variation for the 10 outer shells is presented. Strong shocks developing at the expansion phase are marked with arrows.}
\label{fig:di}
\end{figure}

The initial cycles of model integration are a rather violent phase. Because of the large growth rate of the fundamental mode and imperfect model initialization (scaled F-mode velocity eigenfunction), the rapid growth of pulsation amplitude and irregular pulsation take place. Typically within a few pulsation cycles the parasite overtone modes decay (they are linearly strongly damped) and the amplitude grows steadily till finite amplitude pulsation is reached (typically after few tens -- a few hundred of pulsation cycles). Initial phase of model integration following this scenario is presented in the top panel of Fig.~\ref{fig:di}. For the more luminous and strongly driven models the initial phase is qualitatively different. The development of strong shocks is characteristic for this phase. The outer layer of the model performs significant outward excursions before it contracts and settles on finite amplitude pulsation. This is illustrated in the middle panel of Fig.~\ref{fig:di}. This behaviour is qualitatively the same as observed in the red giant models, see e.g. \cite{lw05}, their fig.~7. As luminosity exceeds a certain value, which mostly depends on the model mass ($L\gtrsim 625\LS$ for $M=0.6\MS$ or $L\gtrsim 1000\LS$ for $M=0.8\MS$) we cannot, in general, calculate the non-linear models -- they brake at the initial phase of the integration. This is illustrated in the bottom panel of Fig.~\ref{fig:di}, in which we also plot the radii variation for the 10 outer model shells. The outward excursion of the outermost layer is too strong; it decouples from the model. This outward excursion is driven by the strong shock that develops during the expansion phase (marked with arrows in Fig.~\ref{fig:di}). Only close to the edges of the IS, in models that are not strongly driven, we can proceed to higher luminosities.

There are several numerical techniques that may be used to pass the violent phases of model integration, but all we have tried in conjunction with our pulsation code fail. Shortening the time step does not help (the default value is already very short). The rapid growth may be suppressed by limiting the pulsation amplitude at the initial phase of model integration. This may be done in several ways, e.g. by setting the eddy viscosity parameter to a high value, and then slowly relaxing it to the desired value during the integration. Still the model brakes, only a bit later. Similarly, using strong artificial viscosity in the outer layers \citep[to treat the shocks,][]{stel75} does not help. We note that \cite{kb88} have given some extra weight to the outermost and penultimate zones to prevent their decoupling during the most extreme hydrodynamic excursions. This is indeed helpful, but only for luminosities below the limits quoted in the previous paragraph. For the same reasons \cite{ow05} apply a non-zero external pressure. This indeed may shift the limiting luminosity to somewhat larger value, but introduces another arbitrary parameter in the calculations, which affects the pulsation dynamics considerably. 

As mentioned above, our calculations are not the first that face the problem of the decoupling of the outer shells. We note that the largest survey of radiative type-II Cepheid models by \cite{kb88} is in fact limited to the same luminosity range -- see their tab.~2. Majority of their models with $M=0.6\MS$ have $L\leq 600\LS$. Only in set H they consider $800\LS$ models with $0.6\MS$, but these are already cool, moderately driven models with very strong artificial viscosity (see their tab.~1). All their models with $M=0.8\MS$ have $L\leq 1000\LS$. As our code and \cite{kb88} code base on the same \cite{stel75} code \citep[see][]{sm08a} we could easily turn the convection off to mimic their computations and to study the radiative models with updated physics. We encounter the instability at exactly the same luminosities as in the convective models. In their introduction, \cite{kb88} review the early hydrodynamic calculations for type-II Cepheids and note that many of the models eject mass due to strong shock waves that develop in their envelopes during the expansion phase. The strong shocks and rapid excursions of the outer stellar layers in low-mass luminous models are  common in the modelling of more luminous AGB and Mira-type stars and are connected to mass loss phenomena \citep[e.g.][]{wood74,tsb79,nakata87}. The radii variation we detect before model brakes is in qualitative agreement with that displayed in the just quoted studies.

The numerous experiments we have done to overpass the violent phase were unsuccessful. We conclude that with the present code, with the fixed-mass Lagrangian mesh, we cannot overpass the violent pulsation phases. A possible fix to the problem, is to allow some mass transfer through the outer bondary. In this case however, the code is no longer Lagrangian, and significant rewriting of the code's scheme is needed. Thus, our model survey is limited in luminosity to $600\LS$ for $0.6\MS$ models and $975\LS$ for $0.8\MS$ models. The exact limits depend on the eddy viscosity parameter and on metallicity. In particular, for higher metallicity models of set A, the luminosity limit is lower. For $M=0.6\MS$ and ${\rm [Fe/H]}=-1.0$ some models with $L\geq 550\LS$ in the center of the IS brake. In the HR diagrams presented in Figs.~\ref{fig:dynamics}, \ref{fig:pc}, \ref{fig:phi21} and \ref{fig:r21} (top left panels) the corresponding area is hatched.

\section{Results}\label{sec:results}
%%%%%%%%%%%%%%%%%%%%%%%%%%%%%%%%%%%%

\subsection{Properties of the static models}\label{ssec:linprop}
%%%%%%%%%%%%%%%%%%%%%%%%%%%%%%%%%%%%%%%%%%%%%%%%%%%%%%%%%%%%%%%%%%%%%%%%%%%

\begin{figure*}
\centering
\resizebox{\hsize}{!}{\includegraphics{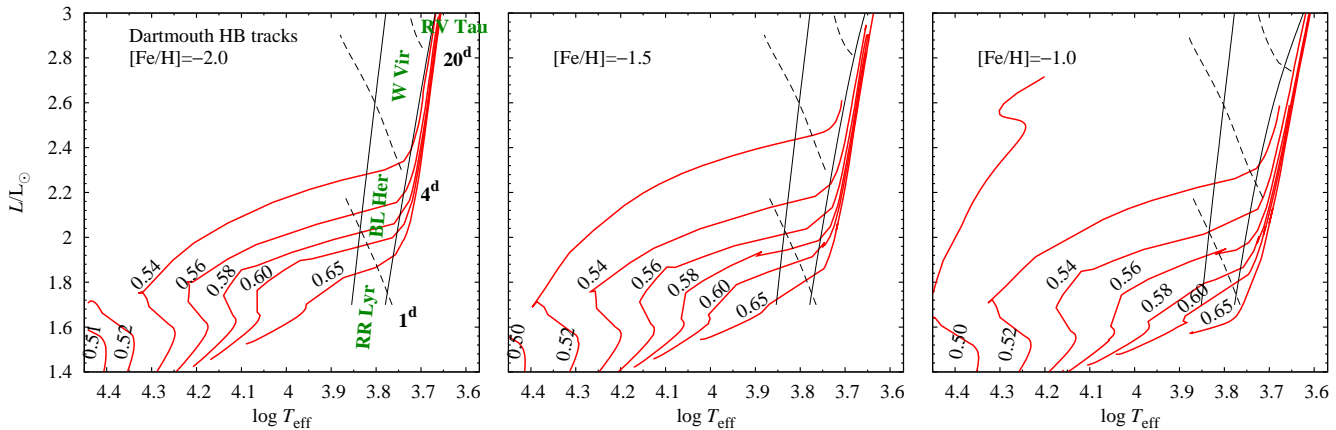}}
\caption{Theoretical HR diagrams for three different metallicities considered in the present study, showing the location of the instability strip and RR~Lyr, BL~Her, W~Vir and RV~Tau domains. Pulsation calculations assume $M=0.6\MS$. Overplotted are horizontal branch tracks from the Dartmouth database \citep{dartmouth}. Each track is labelled with the respective mass value.}
\label{fig:evol}
\end{figure*}

\subsubsection{The stellar evolution context}
In Fig.~\ref{fig:evol} we put our computation domain in the context of stellar evolution theory. The three panels show the location of the instability strip for $0.6\MS$ models of three different metallicities. Over-plotted are stellar evolution tracks for the corresponding metallicities and for several masses (labeled along each track) from the Dartmouth database \citep{dartmouth}. We note that qualitatively the same picture emerges for BaSTI tracks \citep{BaSTI06} \citep[see][]{blherPD}. The lines of constant fundamental mode period of $1$, $4$ and $20$\thinspace d delineate the RR~Lyr, BL~Her, W~Vir and RV~Tau domains. The Figure shows that BL~Her variables should evolve towards the AGB, while their masses should cover $\approx\!0.55\!-\!0.65\MS$ range. The lower the mass the more luminous the track at the crossing with the instability strip. As mass is decreased the tracks become extremely sensitive to the mass. In a narrow mass range, between $0.50\MS$ and $0.54\MS$ (exact value depending on the metallicity) a drastic transition occurs -- low mass tracks evolve towards the blue, never crossing the instability strip. In fact none of the tracks cross the W~Vir domain, except its low luminosity, cool temperature part. Thus, majority of W~Vir variables should be in a different evolutionary stage than BL~Her stars. Unfortunately, in none of the modern evolutionary calculations we have checked\footnote{The following databases were searched: BaSTI \citep{BaSTI06}, PISA \citep{pisa}, Dartmouth \citep{dartmouth}.} we find blue-loops, first computed by \cite{sh70} \citep[see also][]{gingold} that enter the IS at the luminosities corresponding to W~Vir variables. 

Our pulsation calculations do not cover the RV~Tau zone, due to the reasons explained in Sect.~\ref{ssec:inst}. These variables are expected to cross the IS while leaving the AGB. We note however the possibility of other interesting evolutionary scenario for these stars, specifically for those of lower luminosity. The instability strip widens as luminosity is increased. Since its exact width is controlled by the eddy viscosity parameter (and metallicity; in Fig.~\ref{fig:evol}  at $\log L/\LS=3.0$ the width of the IS is equal to $\approx\!1\,350$\thinspace K for ${\rm [Fe/H]}=-2.0$ and to $\approx\!1\,800$\thinspace K for ${\rm [Fe/H]}=-1.0$), it is not unlikely that the stars enter the instability strip while evolving vertically, increasing the luminosity along the AGB.

\subsubsection{The period-luminosity relation}
\begin{figure}
\centering
\resizebox{\hsize}{!}{\includegraphics{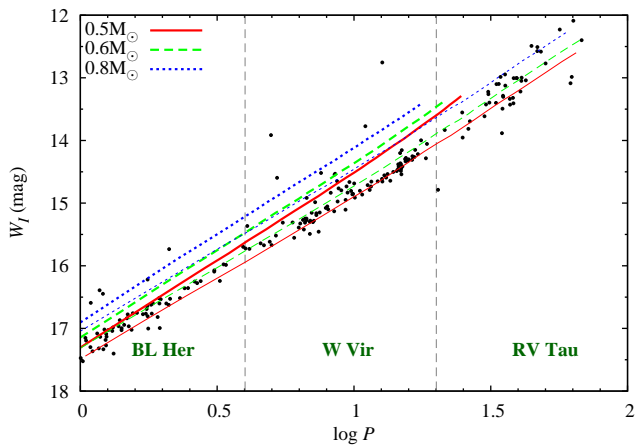}}
\caption{Period-Wesenheit index relation for type-II Cepheids from the LMC \citep[small filled circles,][]{t2cep_lmc} and static models of different masses, $0.5\MS$ (solid lines), $0.6\MS$ (dashed lines) and $0.8\MS$ (dotted lines). For each mass thick and thin lines correspond to the blue and to the red edges of the IS, respectively. Period range of BL~Her, W~Vir and RV~Tau domains is indicated.}
\label{fig:wesenheit}
\end{figure}

Type-II Cepheids follow a distinct period-luminosity relation. Since the observed stars are subject to reddening, in Fig.~\ref{fig:wesenheit} we show the period-Wesenheit index relations computed for our static models and calculated for type-II Cepheids from the LMC \citep{t2cep_lmc} using the two-color OGLE photometry. Reddening free Wesenheit index is defined as $W_I=I-1.55(V-I)$. For the models, $V$ and $I$-band colors were computed using the static \cite{kurucz} atmosphere models\footnote{http://kurucz.harvard.edu/}. LMC variables are used in the comparison as they may be regarded as located at the same distance; we assumed a distance modulus of $18.493$\thinspace mag \citep{lmc_dist}.

The slope of the relations is the same for the models and for the stars. It is clear however, that the lower the mass, the better the agreement between the models and observations. In fact Fig.~\ref{fig:wesenheit} indicates that masses of type-II Cepheids should be low, around $0.5\MS$, or even below, which is in conflict with stellar evolution calculations. We stress however, that the location of the model lines is only approximate and subject to likely large systematic errors. Their source is the use of static equilibrium stellar models and static atmosphere models. Pulsation in type-II Cepheid models is strongly driven. The variables reach large variability amplitudes and the mean structure of the envelope may differ from the static equilibrium structure. This not only causes the non-linear period changes, which will be discussed in Sect.~\ref{ssec:pc}, but also prevents the reliable color determination for the non-linear pulsating model. Quasi-static approach \citep[adopting static atmosphere models and pulsation phase dependent bolometric correction, see][]{sm08a}, which is implemented in our code, is poor already in the case of more gentle pulsations of classical Cepheids or RR~Lyr stars \citep[see e.g.][]{df99}.  

\subsubsection{Linear periods and mode resonances}
Before we discuss the properties of non-linear models, it is important to analyse their linear properties, in particular the linear periods of radial overtones. Although all known type-II Cepheids pulsate in the fundamental mode, and higher order overtones are strongly damped thorough majority of the instability strip (with an important exception of surface modes, see below), the higher order overtones may affect the pulsation properties of the stars through the internal mode resonances. The two well known resonant effects in stellar pulsation are the so-called bump-progressions (Hertzsprung progression) observed in classical Cepheids and caused by the 2:1 resonances \citep[e.g.][Sect.~\ref{ssec:bump}]{ss76,kb89,bmk90,kienzle} and period doubling effect caused by the half-integer resonances \citep[e.g.][Sect.~\ref{ssec:pd}]{mb90}. Half-integer resonances may also cause the modulation of pulsation \citep[e.g.][Sect.~\ref{ssec:mod}]{bk11,blher_mod}. Thus, it is crucial to study whether and which resonances may play a role in type-II Cepheid models. We note that periods we derive are not of asteroseismic precision, since we are using a rather coarse model mesh. Specifically, periods of higher order overtones may differ from that computed with fine-grid and dedicated asteroseismic pulsation codes. Our goal however, is not to make a detailed comparison with the observations, or modelling of the individual stars, but to understand the dynamic properties of the non-linear models. Since these models use exactly the same grid as static equilibrium models, the results of linear analysis provide a firm and self-consistent base for the analysis of non-linear properties of the models. Only in the case of fully consistent linear and non-linear calculations one can infer correlation between non-linear dynamics and mode resonances from linear mode properties \citep{b90}.

\begin{figure}
\centering
\resizebox{\hsize}{!}{\includegraphics{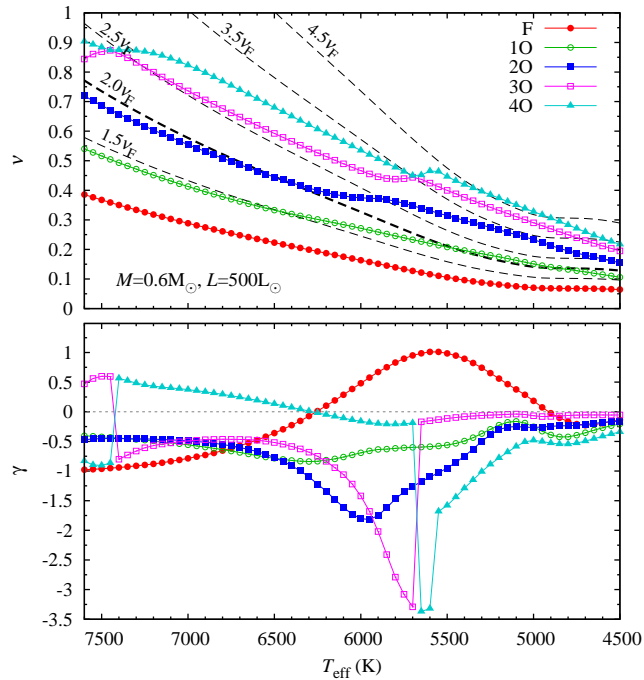}}
\caption{{\it Top:} Radial mode frequencies as a function of model's effective temperature for a constant luminosity sequence. Some integer and half-integer multiples of the fundamental mode frequency are plotted with the dashed lines. {\it Bottom:} Linear growth rates for the modes displayed in the top panel.}
\label{fig:avoidedX}
\end{figure}

In the top panel of Fig.~\ref{fig:avoidedX} we show the variation of mode frequencies with effective temperature in a model sequence of $0.6\MS$ and of constant luminosity, $L=500\LS$ (${\rm [Fe/H]}=-1.5$). The thin dashed lines are placed at $1.5\nu_{\rm F}$, $2.5\nu_{\rm F}$, $3.5\nu_{\rm F}$ and $4.5\nu_{\rm F}$, and indicate the loci o various half-integer resonances between the fundamental mode and higher order overtones. The thick dashed line is placed at $2.0\nu_{\rm F}$ and indicates the loci of the 2:1 resonances. In the bottom panel the corresponding linear growth rates are plotted. The emerging picture is very interesting. The frequencies of the modes do not vary smoothly with the temperature. The lines corresponding to different modes approach each other at some temperatures, just to recede soon after. Avoided crossings between third and fourth overtones are well visible, first crossing in between $7450$\thinspace K and $7400$\thinspace K, the second crossing in between $5700$\thinspace K and $5650$\thinspace K. These two higher order overtones are also unstable on the hot side of the instability strip for the fundamental mode. The phenomenon is not new and is related to so-called strange, or surface modes \citep[e.g.][and references therein]{gb93,bk01}\footnote{There is a controversy regarding the concept of strange modes. According to the original definition, strange mode is a mode that does not have an adiabatic counterpart. Here we discuss modes that have adiabatic counterparts and are trapped in the outer layers. To avoid the confusion, we prefer to call these modes surface (trapped) modes. For the discussion of the controversy see e.g. \cite{dsl05}.}. Their occurrence is related to the narrow hydrogen partial ionization zone which acts as a potential barrier separating the star into two regions. The pulsationally unstable overtone modes are trapped in the outer envelope and are driven entirely in the partial hydrogen ionization zone. The reader is referred to a paper by \cite{byk97} for detailed analysis of the phenomenon, in particular of the origin of avoided crossings. The properties of the surface modes in the case of classical Cepheid and RR~Lyr models were discussed in detail by \cite{bk01}, who found that higher order overtones ($8-12$\thinspace th) are trapped in the outer layers of their models. In the case of type-II Cepheids the phenomenon is observed at lower order overtones ($3-4$\thinspace th), but otherwise it is the same phenomenon. The analysis of \cite{bk01} indicated the plausible existence of variables associated with high order surface modes, the so-called strange Cepheids and strange RR~Lyrae stars, with amplitudes predicted to be in the mmag range. Two plausible Cepheid candidates were identified by \cite{ula}. The same possibility is likely for type-II Cepheids. In this paper however, we focus on the classical instability strip for the fundamental mode and postpone the study of the non-linear properties of the surface modes, except the very interesting beat models addressed in Sect.~\ref{ssec:dm}.

The apparently peculiar run of mode frequencies with effective temperature (Fig.~\ref{fig:avoidedX}) has severe consequence for the location of mode resonances. It is clear that both half-integer and integer resonances may have two loci at a given luminosity level. The scenario changes with the luminosity. As a result, in the HR diagram loci for some resonances are `V' (or `U') shaped. This is illustrated in a few figures in this paper in which the resonance loci are plotted, in particular in Figs.~\ref{fig:rezhi} and \ref{fig:rez21}. We note that resonances may affect the pulsation properties of the models not only exactly at the resonance center, determined from linear periods, but also close to it, in the {\it resonance band} along its loci. Within the band, non-linear phase synchronisation occurs. As an example, the Hertzsprung bump progressions, caused by the 2:1 resonance between the fundamental mode and the second overtone, affects the shape of light curves of classical Cepheids in a wide period range. Due to non-linear phase synchronization the second overtone is not detected as an independent frequency however (since $\nu_{\rm 2O}=2\nu_{\rm F}$), but manifests as a distortion of the light curve -- a bump -- position of which depends on the linear period ratio \citep[e.g.][]{bmk90}. In the context of the present models other effect, the non-linear period change, is also crucial (Sect.~\ref{ssec:pc}). The period changes are large for the luminous models and can significantly shift the resonance positions in the HR diagram. The amount and direction of the shift are hard to predict, as we can only determine the non-linear period of the fundamental mode, and not of the higher order overtones. The resonant effects are discussed in more detail in Sect.~\ref{ssec:pd} (half-integer resonances) and \ref{ssec:bump} (2:1 resonances).

\subsection{Analysis of non-linear models}
%%%%%%%%%%%%%%%%%%%%%%%%%%%%%%%%%%%%%%%%%%

\begin{figure*}
\centering
\resizebox{\hsize}{!}{\includegraphics{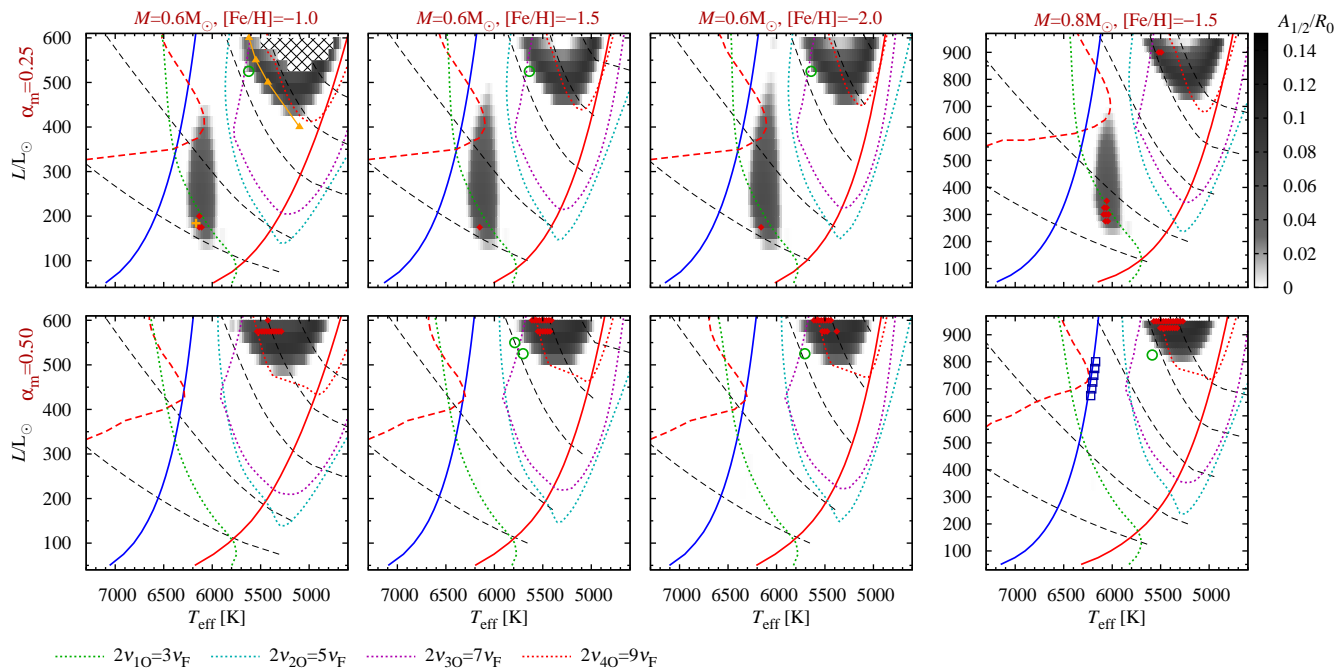}}
\caption{Dynamical scenario for the computed models. White areas within the instability strip (limited by the thick blue and red lines -- edges of the instability strip) correspond to single-periodic fundamental mode pulsation. Grey-shaded areas correspond to period doubling domains. Darker the area, larger the amplitude of alternations (the grey scale codes the value of $A_{1/2}/R_0$, see eq.~\eqref{eq:fsum}, as indicated in the top right part of the figure). Diamonds within period doubling domains point the models in which period-4 pulsation was detected. Open circles indicate the models in which periodic modulation of pulsation was detected. Squares are double-periodic F+4O models. Loci of some half-integer resonances are marked with the dotted lines. Dashed lines are lines of constant fundamental mode period (linear value) equal to (from bottom to top): 2, 4, 8, 12 and 16\thinspace d. In the top left panel, orange triangles connected with a solid line indicate the edge of the period doubling domain in the calculations of Kov\'acs \& Buchler (1988). A plus symbol in the same panel indicates the location of the best model for T2CEP-279 with $0.6\MS$ computed by Smolec et al. (2012).}
\label{fig:dynamics}
\end{figure*}

For all models the time series of radius variation from the last $100$ cycles of model integration was analysed. First, discrete Fourier transform of time series was computed in order to identify the dominant period. Then the data was fitted with the sine series of the following form:
\begin{align}
R(t) = R_0  + &\sum_{k=1}^N A_k \sin\big(2\uppi k \nu t + \phi_k\big)+\nonumber \\
+& \sum_{k=0}^K A_{(2k+1)/2}\sin\bigg[2\uppi\frac{2k+1}{2}\nu t + \phi_{(2k+1)/2}\bigg]+\nonumber \\
+& \sum_{k=0}^L A_{(2k+1)/4}\sin\bigg[2\uppi\frac{2k+1}{4}\nu t + \phi_{(2k+1)/4}\bigg].
\label{eq:fsum}
\end{align}
The first line above represents a well known $N$-th order Fourier series, a most common representation of the non-linear periodic signals. $\nu$ is the dominant frequency, which corresponds to the fundamental mode in the vast majority of the models. In the second line, sub-harmonic series is present, i.e. the frequencies in the consecutive sine terms correspond to $\nu/2$ and its harmonics. The sub-harmonic series is necessary to describe the period doubling phenomenon. In the third line, the terms necessary to model the period-4 pulsations are given. Since no further period doublings were detected, there was no need to supplement the formula with further terms. The amplitudes and phases of the harmonic series were used to construct the Fourier decomposition parameters \citep{sl81}, amplitude ratios, $R_{k1}=A_k/A_1$, and phase differences, $\varphi_{k1}=\phi_k-k\phi_1$. The time-series for each model was folded with the dominant period and the resulting radius variation curves for all the models were inspected by eye. Five distinct scenarios were observed (see Fig.~\ref{fig:dynamics}): (i) single-periodic fundamental mode pulsation (white areas within IS in Fig.~\ref{fig:dynamics}); (ii) period-doubled pulsation (grey-shaded domains in Fig.~\ref{fig:dynamics}); (iii) period-4 pulsation, always within period doubling domains (diamonds in Fig.~\ref{fig:dynamics}); (iv) periodic modulation of pulsation (open circles in Fig.~\ref{fig:dynamics}); (v) double-periodic (beat) pulsation in the fundamental mode and in the fourth radial overtone (squares in Fig.~\ref{fig:dynamics}). These interesting forms of pulsations are discussed in the following Sections.

We also note that in some of the models single-periodic, first overtone pulsation is detected (after a mode-switch during the model integration). This behaviour was noticed only for the models of set A, of lowest luminosities and close to the blue edge of the fundamental mode instability strip. These are in fact short period ($P\!<\!0.5$\thinspace d) first overtone RR~Lyr-type models.

\subsection{Non-linear period changes}\label{ssec:pc}
%%%%%%%%%%%%%%%%%%%%%%%%%%%%%%%%%%%%%%%%%%%%%%%%%%%%%

\begin{figure*}
\centering
\resizebox{\hsize}{!}{\includegraphics{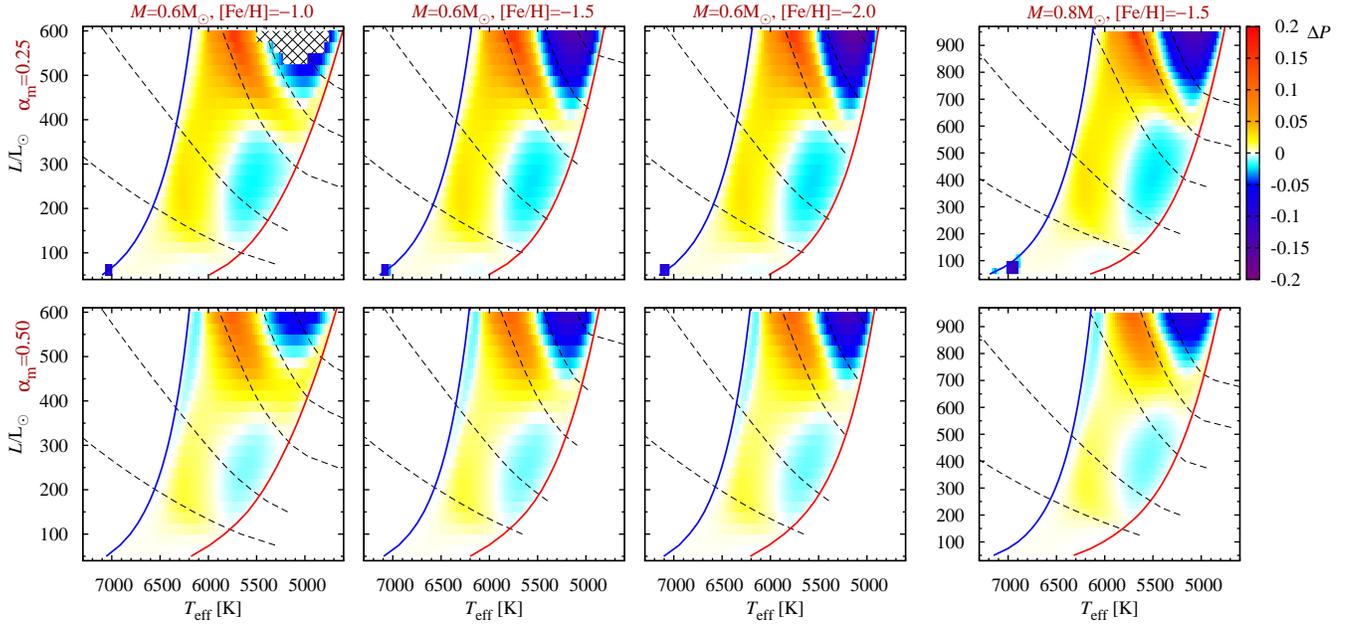}}
\caption{Non-linear period changes for the computed models. Colors code the relative difference between the non-linear and linear period of the model, i.e., $\Delta P=(P_{\rm nl}-P_{\rm l})/P_{\rm l}$, as indicated in the top-right part of the Figure. Note that the significant period shortening, visible close to the blue edge of the IS at the lowest luminosities of $\alpha_{\rm m}=0.25$ models, corresponds to RR~Lyr-type models that switched into first overtone pulsation. The cyan-blue domains correspond to period shortening, while yellow-red domains correspond to period lengthening.}
\label{fig:pc}
\end{figure*}

In Fig.~\ref{fig:pc} we show the relative change of the pulsation period with respect to the linear value, $\Delta P=(P_{\rm nl}-P_{\rm l})/P_{\rm l}$ ($P_{\rm nl}$ and $P_{\rm l}$ are non-linear and linear period, respectively). The detected pattern is quite complex. We first focus on models of set A. The most striking feature is the decrease of non-linear periods with respect to the linear ones in some regions of the HR diagram. In the case of classical Cepheid models or RR~Lyr models, the non-linear period is typically longer than the linear period by up to two per cent. Type-II Cepheid models are different, however. They are strongly non-adiabatic, with large amplitude variation. They are more similar to red giant or Mira-type models, in which period shortening is also observed \citep[see e.g.][]{w07}. The effect was described e.g. in \cite{yt96} and interpreted as due to significant rearrangement of stellar structure caused by the large amplitude pulsation. As a result the mean structure of the pulsating envelope differs from the static structure. In the models of red giants period changes in both directions are found, but no systematic study of the effect was carried \citep{w07}.

In Fig.~\ref{fig:pc} we observe that the brighter the models, the more pronounced the period change is. It may be as high as $\pm 15$\thinspace per cent. For models of set A there are two domains at which non-liner period is shorter than the linear one. Both are adjacent to the red edge of the IS; period shortening is much more pronounced in the more luminous domain. For models of set B the picture is qualitatively the same, except that another domain with period shortening appears. It is thin, adjacent to the blue edge of the instability strip, and period shortening is not as significant there (up to $2$\thinspace per cent).

\begin{figure}
\centering
\resizebox{\hsize}{!}{\includegraphics{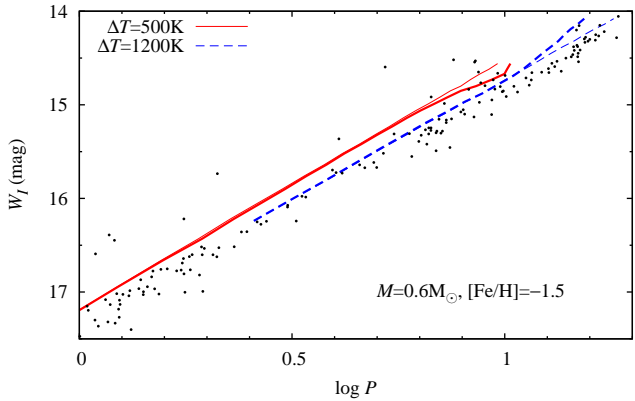}}
\caption{The effect of non-linear period change on the period-Wesenheit index relation for type-II Cepheids. Thick and thin lines correspond to non-linear and liner periods, respectively, for model sequences running at a distance of $500$\thinspace K (solid) and $1200$\thinspace K (dashed) from the blue edge of the IS.}
\label{fig:wesenheit2}
\end{figure}

It is interesting to check whether the described period changes can have significant effect on the period--Wesenheit index relations. It is the case for Mira variables, see e.g. \cite{lw05}. In Fig.~\ref{fig:wesenheit2} we plot the period-Wesenheit index relations for models of set A, with $M=0.6\MS$ and ${\rm [Fe/H]}=-1.5$. The models either run at a distance of $500$\thinspace K from the blue edge of the IS (solid lines), and at high luminosities cross the domain in which period lengthening is most significant, or at a distance of $1200$\thinspace K from the blue edge (dashed lines), and cross the domain at which period shortening is most significant. Thin lines correspond to linear periods, while thick lines correspond to non-linear periods. The Wesenheit index in both cases was computed assuming static values of luminosity and effective temperature. The thick and thin lines mostly overlap. Only for longer periods there is a noticeable, but small difference. We conclude that period changes do not affect significantly the period-Wesenheit index relations, at least in the parameter range covered by our models.

\subsection{Dynamical phenomena -- period doubling domains}\label{ssec:pd}
%%%%%%%%%%%%%%%%%%%%%%%%%%%%%%%%%%%%%%%%%%%%%%%%%%%%%%%%%%%%%%%%%%%%%%%%%%

The period doubling domains are marked with grey-shaded areas in Fig.~\ref{fig:dynamics}; the darker the area, the more pronounced the alternations. The grey scale represents the relative amplitude of alternations, $A_{1/2}/R_0$, eq.~\eqref{eq:fsum}. We first focus on $0.6\MS$ models of set A, scenario is qualitatively the same for $0.8\MS$ models, except that the period doubling domains are shifted towards the higher luminosities. There are two distinct period doubling domains for models of set A. The first domain runs nearly vertically, between luminosities of $\approx\!150\LS$ and $\approx\!450\LS$. It is centered at $\teff\!\approx\!6100-6150$\thinspace K. A weak dependence on metallicity is apparent; in the more metal poor models the domain is slightly larger, extends towards slightly higher luminosities and covers a slightly larger effective temperature range. Also, amplitude of the alternations is somewhat higher for metal poor models. Within the domain, fundamental mode periods cover a range from $\approx\!2$\thinspace d up to $\approx\!6.5$\thinspace d for the most luminous models. In Tab.~\ref{tab:pdprop} we characterize the extent of this period doubling domain more quantitatively, providing minimum and maximum values of the effective temperature, luminosity and period. A sample radius variation curves, showing the period doubling effect, are plotted in Fig.~\ref{fig:pdcurves}. These are models of $0.6\MS$ and ${\rm [Fe/H]}=-1.5$, running at a constant distance of $500$\thinspace K from the blue edge of the instability strip. Models at the bottom and at the top of the figure are single-periodic pulsators. Interestingly, the $175\LS$ model shows period-4 pulsation, to be discussed in subsequent paragraphs.

\begin{table*}
\begin{tabular}{lllrrrrr}
    &         &        & \multicolumn{3}{c}{low $L$ domain} & \multicolumn{2}{c}{high $L$ domain} \\
set & $M/\MS$ & [Fe/H] & $T_{\rm min}$/$T_{\rm max}$ & $L_{\rm min}$/$L_{\rm max}$ & $P_{\rm min}$/$P_{\rm max}$ & $L_{\rm min}$ & $P_{\rm min}$ \\
    &         &        & (K)                     &  $(\LS)$                &  (d)                    & $(\LS)$     &  (d)\\ 
\hline
A   & $0.6$   & $-1.0$ & 5893/6248 & 150/425 & 2.08/5.32 & 450 & 10.12\\
A   & $0.6$   & $-1.5$ & 5885/6280 & 150/475 & 2.04/5.83 & 475 &  9.91\\
A   & $0.6$   & $-2.0$ & 5894/6273 & 150/500 & 2.02/6.16 & 475 &  9.86\\
A   & $0.8$   & $-1.5$ & 5831/6181 & 250/675 & 2.67/6.55 & 750 & 12.65\\
B   & $0.6$   & $-1.0$ & $-$       & $-$     & $-$       & 500 &  9.53\\
B   & $0.6$   & $-1.5$ & $-$       & $-$     & $-$       & 525 &  9.79\\
B   & $0.6$   & $-2.0$ & $-$       & $-$     & $-$       & 525 &  9.73\\
B   & $0.8$   & $-1.5$ & $-$       & $-$     & $-$       & 825 & 12.56\\
\hline
\end{tabular}
\caption{Extent of the two period doubling domains (of lower and of higher luminosity) detected in the models -- cf. Fig.~\ref{fig:dynamics}.}\label{tab:pdprop}
\end{table*}

The second period doubling domain extends at higher luminosities and is present in model grids adopting both sets of convective parameters. This domain has a `V' shape: at highest considered luminosities there are two period doubling domains centered at different effective temperatures, which then merge into one, as luminosity is decreased. The effect is not marked for models of set B. The location of the cool edge of this domain is dictated by the location of the red edge of the IS. Pulsation periods within this domain start from $\approx 9.5$\thinspace d (see Tab.~\ref{tab:pdprop}). Unfortunately, the full extent of this domain, and its shape at even higher luminosities, cannot be studied due to reasons outlined in Sect.~\ref{ssec:inst}. In Tab.~\ref{tab:pdprop} we just provide the values of minimum luminosity and period for the computed model sequences. 

In models of set B the eddy-viscous dissipation is increased, $\alpha_{\rm m}=0.5$. This should lead to the shrinking of the period doubling domains, see fig.~15 in \cite{blherPD} and also \cite{kms11}. Indeed, in the bottom panels of Fig.~\ref{fig:dynamics} we observe that the low luminosity period doubling domain is not present at all, while the high luminosity domain has shrunk considerably. Interestingly, period-4 cycle models appear within this domain, which were not present in the models of set A, except one model of $M=0.8\MS$.

What is the origin of the period doubled pulsation? From dynamical point of view, period doubling bifurcation occurs as control parameter, e.g. temperature in a model sequence of constant luminosity, varies \citep[e.g.][]{seydel}. The bifurcation may be triggered by the half-integer resonance between the pulsation modes, as first analysed by \cite{mb90}. The only reliable tools that allow a firm connection of period doubling bifurcation with particular resonance are ({\it i}) the relaxation technique \citep{stel74}, which allows to converge the model to strict limit cycle pulsation and the following ({\it ii}) Floquet analysis, which allows to investigate the stability of the limit cycle. Details of the method are provided in \cite{mb90}.  Analysis of the Floquet coefficient not only allows to point when the limit cycle becomes unstable, but also clearly indicate which pulsation mode is responsible for its destabilization. The analysis is not that straightforward, however: for type-II Cepheid models Floquet coefficients strongly vary along a model sequence and dense model grid is necessary to trace their evolution. Based on the Floquet analysis \cite {mb90} connected the period doubling bifurcation observed in the sequence of models of \cite{kb88} (leading to chaos through the period doubling cascade) to the 5:2 resonance between the fundamental mode and the second overtone.

Unfortunately, this technique is not implemented in our convective codes. We may only indicate which resonance may play a role based on the correlation between liner period ratios and model dynamics. This is however tough, as already noted by \cite{mb90} due to non-linear period shifts, which may be particularly strong in luminous type-II Cepheid models (Sect.~\ref{ssec:pc}, Fig.~\ref{fig:pc}).

\begin{figure*}
\centering
\resizebox{\hsize}{!}{\includegraphics{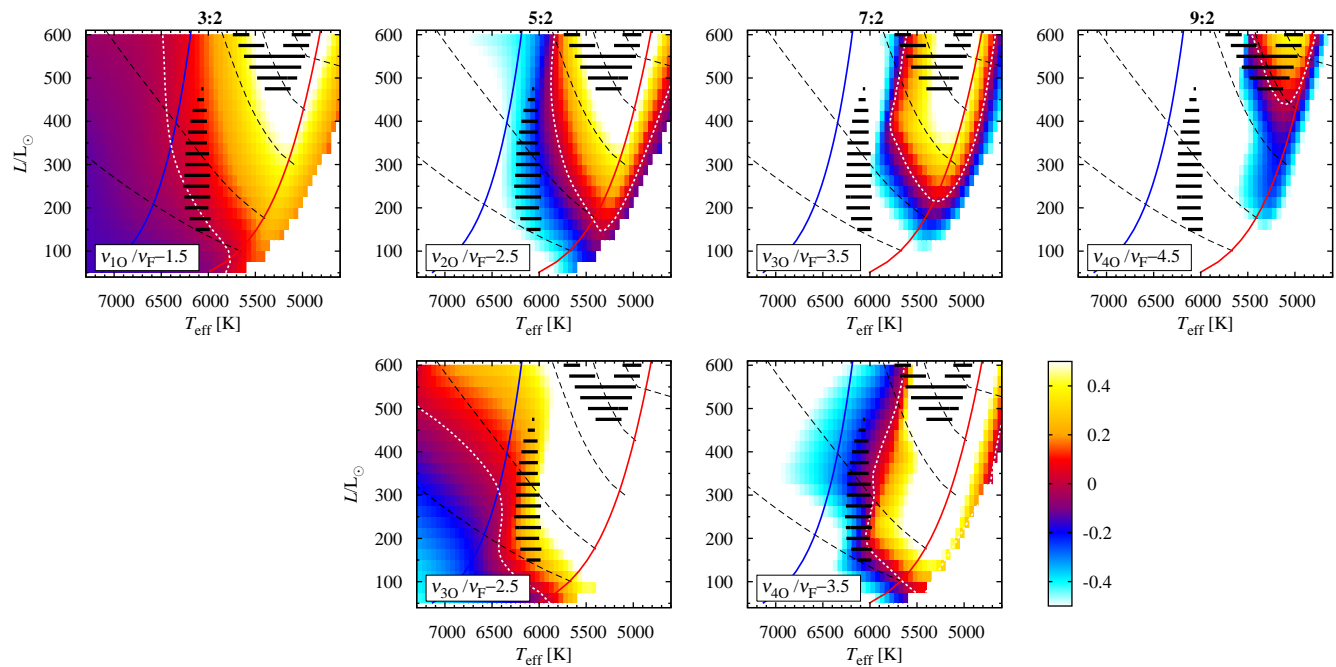}}
\caption{Theoretical HR diagrams with loci of several half-integer resonances involving fundamental mode and higher order overtones of the (from left to right) 3:2, 5:2, 7:2 and 9:2 type. Colors code the relevant resonance mismatch parameters defined in the bottom-left part of each panel. Edges of the IS are also plotted, as well as lines of constant fundamental mode period (dashed lines). In addition the domains of period doubling are marked with thick horizontal line segments.}
\label{fig:rezhi}
\end{figure*}

\begin{figure}
\centering
\resizebox{\hsize}{!}{\includegraphics{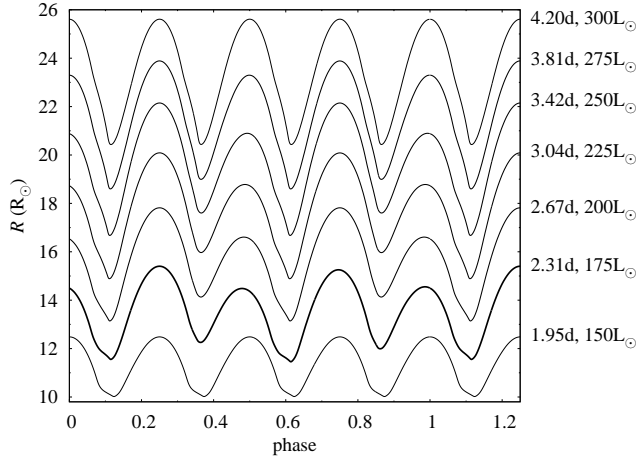}}
\caption{Radius variation curves along a model sequence ($M=0.6\MS, {\rm [Fe/H]}=-1.5$) running at a distance of $500$\thinspace K from the blue edge of the IS and crossing the low luminosity period doubling domain. Pulsation period (undoubled) and luminosity for each curve are given on the right-hand side of the figure. For clarity, for each model brighter than $150\LS$, the curve is shifted up by $1\RS$ with respect to the lower luminosity curve.}
\label{fig:pdcurves}
\end{figure}

For the low luminosity period doubling domain the situation is rather clear. This domain is not new. It was first revealed in the radiative calculations of \cite{bm92}, and led the authors to the prediction that period doubling phenomenon should occur in BL~Her stars. Indeed, twenty years later the first BL~Her star with period doubling, OGLE-BLG-T2CEP-279, was discovered \citep[][T2CEP-279 in the following]{blher_ogle}. The analysis of \cite{blherPD} confirmed the existence of period doubling domain, but, since that paper was focused on modelling the single star, its extent was not determined; models of \cite{blherPD} fall along a line of constant period, equal to the period of T2CEP-279 ($\approx\!2.4$\thinspace d). Location of the best $0.6\MS$ model for T2CEP-279 is marked with cross in the top-left panel in Fig.~\ref{fig:dynamics}. Using the Floquet analysis, \cite{bm92} found that the 3:2 resonance between the fundamental mode and the first overtone, $2\nu_{\rm 1O}=3\nu_{\rm F}$, is responsible for the occurrence of period doubling in their models. The loci of this resonance is marked with green dotted line in Fig.~\ref{fig:dynamics} and indeed crosses the discussed period doubling domain. In Fig.~\ref{fig:rezhi} we plot the loci of several half-integer resonances that may be relevant in the context of period doubled pulsation. In all panels models with $0.6\MS$ and ${\rm [Fe/H]}=-1.5$ of set A are plotted; results are similar for other model sequences. For each considered half-integer resonance, $2\nu_k=(2n+1)\nu_0$, we define a mismatch parameter as $\Delta=\nu_k/\nu_0-(2n+1)/2$ and in Fig.~\ref{fig:rezhi} color-code the mismatch value. We also mark the extent of the two period doubling domains with thick horizontal line segments. We note that only for the 3:2 resonance with the first overtone, $2\nu_{\rm 1O}=3\nu_{\rm F}$, the mismatch parameter is small ($\Delta<0.1$) within the entire discussed period doubling domain. It is not the case for other half-integer resonances, in particular for the 5:2 (with the third overtone) and the 7:2 (with the fourth overtone) resonances which also run close to the discussed domain. The same conclusion was reached by \cite{blherPD} who analysed the potential role of other half-integer resonances for their $2.4$\thinspace d model sequence (which falls within the domain discussed here). We also note that within this domain the non-linear shift of the resonance is likely not strong as non-linear period change for the fundamental mode is small (always below 3\thinspace per cent, Fig.~\ref{fig:pc}). 

Which half-integer resonance may be responsible for the high luminosity period doubling domain? Interestingly, the loci of the 5:2, 7:2 and 9:2 resonances with the second, third, and fourth overtones respectively, also follow  a `V' shape, as noted in Sect.~\ref{ssec:linprop} and as is visible in the top panels of Fig.~\ref{fig:rezhi}. Unfortunately, we cannot firmly point which resonance (or resonances) are crucial. The analysis is hampered by strong non-linear period shifts in this domain, by up to $15$\thinspace per cent in both directions (Fig.~\ref{fig:pc}). Comparing Figs.~\ref{fig:dynamics} and \ref{fig:pc} we note that within the gross part of the period doubling domain the non-linear period of the fundamental mode is significantly shorter, but at the hot side of the period doubling domain the abrupt transition between significant period decrease and significant period increase takes place. The periods of the overtone modes are likely strongly affected as well. Therefore, based on resonance loci calculated using linear periods, we may only conclude that each of the three resonances, 5:2 (with the second overtone), 7:2 (with the third overtone) and 9:2 (with the fourth overtone), may play a role in driving the period doubled pulsation. The 7:2 resonance with the fourth overtone is likely too far to play a role. 

Although we cannot firmly point which resonance is responsible for the period doubling bifurcation in the models, we can argue in favour of the 5:2 resonance with the second overtone, $2\nu_{\rm 2O}=5\nu_{\rm F}$. First, we note that as the resonance order increases, the band in the HR diagram in which the mismatch parameter is small (reddish area in Fig.~\ref{fig:rezhi}) clearly decreases. For the 5:2 resonance it is still relatively large and we may expect that this resonance can affect the pulsation over significantly larger area in the HR diagram than the 7:2 or 9:2 resonances. Second, the discussed period doubling domain is likely not new, but is the same domain as found by \cite{kb88}. The one-to-one comparison of their models with ours is not possible for obvious reasons: their models are radiative and use old opacity data. Still we can trace the appearance of the period doubling in their models in the HR diagram. From their tab.2 we picked the model sequences of $0.6\MS$ and $Z=0.005$, i.e. sequences C, D, E and G. Then, in the top-right panel of Fig.~\ref{fig:dynamics} (${\rm [Fe/H]}=-1.0$ seems most suitable for comparison) we mark and connect the models in which period doubling was observed for the first time (filled triangles). The agreement between the location and shape of the hot boundary of the period doubling domain in our calculations and in calculations of \cite{kb88} is striking. Likely, these are the same period doubling-domains, just as in the case of low luminosity period doubling domain and \cite{bm92} models. Based on the Floquet analysis, \cite{mb90} showed that the 5:2 resonance with the second overtone triggered the period doubling bifurcation in \cite{kb88} models. Therefore, we point that the 5:2 resonance is also most likely crucial in the present convective models, but we admit that firm evidence is lacking.

Except that the period doubling bifurcation occurs at roughly the same place in the HR diagram, our models and models of\cite{kb88} display qualitatively different behaviour. As effective temperature was decreased, \cite{kb88} detected a period doubling cascade en route to chaos, which is not present in our models. Instead, for some high luminosity model sequences we detect only one subsequent period doubling bifurcation, which leads to period-4 pulsation domain within period doubling domain (diamonds in Fig.~\ref{fig:dynamics}). No further bifurcations are detected in our models. As temperature is decreased, we observe the transition from period-4 pulsation to period doubled pulsation, and finally to single-periodic pulsation. We note that pulsation dynamics is strongly sensitive to the amount of viscous dissipation in the models, which is controlled through artificial viscosity parameters in the radiative models and eddy viscosity parameter in the convective models. The larger the eddy viscosity the simpler the dynamics, which is clear already from Fig.~\ref{fig:dynamics}. Also, the lower the model non-linearity the simpler the dynamics. Consequently, the large value of eddy viscosity in our models and the presence of the red edge of the IS (not present in the radiative models) suppresses the further period doublings and the appearance of chaos. If eddy viscosity is strongly reduced, complex dynamics appears in the models \citep{blher_mod,blher_chaos}

For all the mentioned resonances, 5:2, 7:2 and 9:2, their bands are small as compared to the 3:2 resonance, and across the vast part of the discussed period doubling domain the respective mismatch parameters must be large ($\Delta>0.2$, Fig.~\ref{fig:rezhi}). It is the case also for the favoured 5:2 resonance. We may speculate that the half-integer resonance only triggers the period doubling bifurcation and play no essential role across the entire period doubling domain. 

\subsection{Dynamical phenomena -- modulation of pulsation}\label{ssec:mod}
%%%%%%%%%%%%%%%%%%%%%%%%%%%%%%%%%%%%%%%%%%%%%%%%%%%%%%%%%%%%%%%%%%%%%%%%%%%

\begin{figure}
\centering
\resizebox{\hsize}{!}{\includegraphics{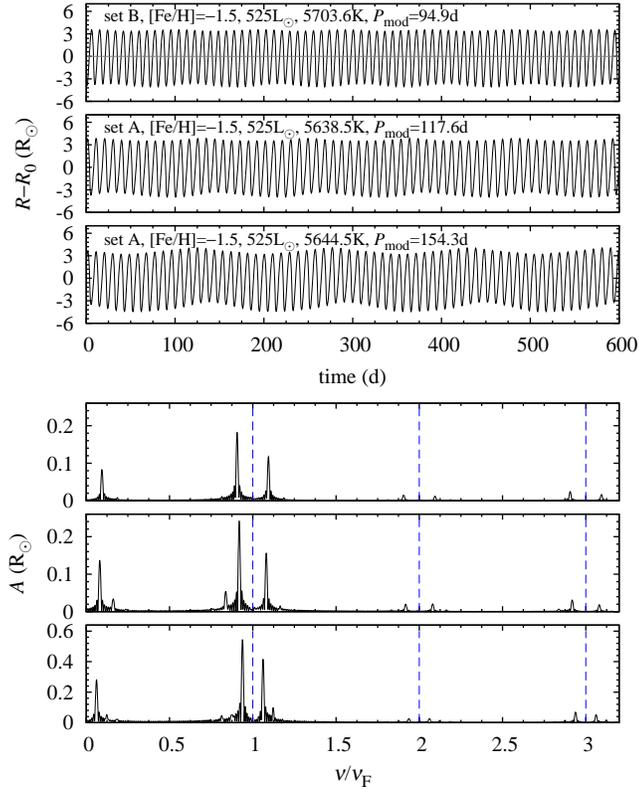}}
\caption{Illustration of modulation detected in some of the models. In the top panels radius variation for three exemplary models is plotted. In the bottom panels frequency spectrum (after prewhitening with the fundamental mode frequency and its harmonics; dashed lines) for these models is plotted.}
\label{fig:mod}
\end{figure}

Altogether in 7 models of our survey we detect periodic modulation of pulsation. These models are marked with open circles in Fig.~\ref{fig:dynamics}. In a grid of models of given mass and metallicity we detect only one or two (set B, $0.6\MS$, ${\rm [Fe/H]}=-1.5$) modulated models. Interestingly, all modulated models we have computed, are located nearly exactly at the same place in the HR diagram, close to the high luminosity period doubling domain. Are these models singular oddballs? No doubt the models show the pronounced modulation of pulsation -- examples are presented in the top panels of Fig.~\ref{fig:mod}. The presence of only one or two models with modulation of pulsation in our relatively dense model grids indicates that the domains with modulation of pulsation must be very small. To investigate the problem we have computed additional models of set A, with $M=0.6\MS$ and ${\rm [Fe/H]}=-1.5$, along a constant luminosity line crossing the modulated model from Fig.~\ref{fig:dynamics} ($L=525\LS$) with a finer spacing of only $1$\thinspace K in effective temperature. We clearly detect a domain in which pulsation is modulated, the models in between $5647.5$\thinspace K and $5637.5$\thinspace K are all modulated. The domain is thus very narrow, $\sim\!10$\thinspace K-wide in effective temperature.

In Fig.~\ref{fig:mod} (top) we show the time series of radius variation for three exemplary models. Modulation is clear; it is also clear that the mean radius is strongly modulated. This is further confirmed with analysis of the frequency spectra, plotted in the bottom panels of Fig.~\ref{fig:mod} for the same models. Dashed lines mark the location of the prewhitened fundamental mode frequency and its harmonics. Equidistant modulation side-peaks are well visible. In all models the lower frequency side-peak at the fundamental mode frequency, $A_-$, has the highest amplitude. After prewhitening with the triplet components, further multiplet components are detected (quintuplets are already visible in Fig.~\ref{fig:mod}). Also a signal at modulation frequency (and at its harmonics) is present and is strong, in agreement with the apparent modulation of the mean radius. 

Tab.~\ref{tab:mod} summarizes the basic properties of the modulated models, their luminosity, effective temperature, distance from the blue edge of the IS, pulsation period and modulation period, amplitude ratio between the amplitude of the highest modulation side-peak and amplitude of the fundamental mode, $A_-/A_{\rm F}$, and finally the asymmetry parameter, constructed using the amplitudes of the lower and higher frequency triplet components, $A_-$ and $A_+$; $Q=(A_+-A_-)/(A_++A_-)$ \citep{alcock_rrab}. All models are qualitatively similar. Modulation periods are a few up to several times longer than pulsation periods. Frequency spectra are also qualitatively the same for all the models. For the $10$-K wide modulation domain, for which detailed calculations were conducted, we note a steep increase of modulation amplitude at the hot edge of the domain (the first model has a relative amplitude of 13\thinspace per cent, while no modulation was detected in a model hotter by $1$\thinspace K) and strong variation of the modulation period as a function of effective temperature.

It is not the first time modulation is detected in type-II Cepheid models. Our previous paper \citep{blher_mod} was entirely devoted to the study of modulation of pulsation detected in BL~Her-type models computed with the strongly decreased eddy viscosity parameter ($\alpha_{\rm m}=0.05$). In that study, modulation was detected on top of pronounced period doubling effect. The mechanism behind such dynamics is a half-integer resonance between the pulsation modes. The mechanism was proposed by \cite{bk11} to explain the Blazhko effect in RR~Lyrae stars, after period doubling in Blazhko variables was discovered in {\it Kepler} observations of these stars \citep{kol10,szaboPD}. Using the amplitude equation formalism, \cite{bk11} showed, that the same half-integer resonance that most likely underlies the period doubling effect in modulated RR~Lyr variables \citep[a 9:2 resonance between the fundamental mode and the ninth overtone, which is a surface mode,][]{kms11} can also cause the modulation of pulsation. Unfortunately, the RR~Lyr-type hydrodynamic models do not show the modulation of pulsation, only the period doubling effect is detected in the models \citep{kms11, sm15epj}. The resonant mechanism is clearly operational in the case of BL~Her models of \cite{blher_mod} but different half-integer resonance is in action, the 3:2 resonance between the fundamental mode and the first overtone. What causes the modulation in the present, W~Vir-type models? The models are located close, but outside the high luminosity period doubling domain. They show the modulation of pulsation only, not the alternations. Still, it is clear from the analysis of amplitude equations, that half-integer resonances may drive the modulation. Just as in the case of high luminosity period doubling domain we cannot firmly point which half-integer resonance might be in action, due to the strong period changes at higher luminosities.

\begin{table*}
\begin{tabular}{lllrrrrrr}
set & $M/\MS$ & [Fe/H] & $L/\LS$ & $\teff$ ($\Delta T$) (K) & $P_{\rm F}$ (d) & $P_{\rm mod}$ (d) & $A_-/A_{\rm F}$ & $Q$ \\
\hline
A   & $0.6$   &  $-1.0$  & $525$  & $5623.4$ (600) &  $9.735$ & $116.28$  &  $0.061$ & $-0.22$ \\ 

A   & $0.6$   &  $-1.5$  & $525$  & $5647.5$ (591) &  $9.471$ & $191.32$  &  $0.130$ & $-0.14$ \\
A   & $0.6$   &  $-1.5$  & $525$  & $5646.5$ (592) &  $9.479$ & $173.22$  &  $0.142$ & $-0.14$ \\
A   & $0.6$   &  $-1.5$  & $525$  & $5645.5$ (593) &  $9.486$ & $161.29$  &  $0.148$ & $-0.15$ \\
A   & $0.6$   &  $-1.5$  & $525$  & $5644.5$ (594) &  $9.494$ & $154.31$  &  $0.148$ & $-0.14$ \\
A   & $0.6$   &  $-1.5$  & $525$  & $5643.5$ (595) &  $9.503$ & $149.30$  &  $0.144$ & $-0.14$ \\
A   & $0.6$   &  $-1.5$  & $525$  & $5642.5$ (596) &  $9.511$ & $144.61$  &  $0.139$ & $-0.14$ \\
A   & $0.6$   &  $-1.5$  & $525$  & $5641.5$ (597) &  $9.520$ & $139.78$  &  $0.130$ & $-0.15$ \\
A   & $0.6$   &  $-1.5$  & $525$  & $5640.5$ (598) &  $9.528$ & $133.28$  &  $0.117$ & $-0.17$ \\
A   & $0.6$   &  $-1.5$  & $525$  & $5639.5$ (599) &  $9.533$ & $124.39$  &  $0.095$ & $-0.19$ \\
A   & $0.6$   &  $-1.5$  & $525$  & $5638.5$ (600) &  $9.538$ & $117.59$  &  $0.067$ & $-0.22$ \\
A   & $0.6$   &  $-1.5$  & $525$  & $5637.5$ (601) &  $9.543$ & $122.04$ &  $0.002$ & $-0.22$ \\

A   & $0.6$   &  $-2.0$  & $525$  & $5642.8$ (600) &  $9.472$ & $145.19$  &  $0.111$ & $-0.15$ \\
B   & $0.6$   &  $-1.5$  & $525$  & $5703.6$ (550) &  $8.926$ &  $94.89$  &  $0.050$ & $-0.22$ \\
B   & $0.6$   &  $-1.5$  & $550$  & $5788.9$ (450) &  $8.683$ &  $68.97$  &  $0.014$ & $-0.20$ \\
B   & $0.6$   &  $-2.0$  & $525$  & $5707.1$ (550) &  $8.875$ &  $99.74$  &  $0.054$ & $-0.20$ \\
B   & $0.8$   &  $-1.5$  & $825$  & $5587.0$ (600) & $11.743$ & $132.29$  &  $0.084$ & $-0.22$ \\
\hline
 %Q=(A+ - A-)/(A+ + A-)
\end{tabular}
\caption{Properties of the modulated models, including additional models of set A ($M=0.6\MS$ and ${\rm [Fe/H]}=-1.5$) computed with $1$\thinspace K-step in effective temperature. Except mass and metallicity, luminosity and effective temperature of the models is given (including the distance from the blue edge of the IS, $\Delta T$), pulsation and modulation periods, amplitude ratio between the highest modulation side peak and amplitude of the fundamental mode, $A_-/A_{\rm F}$, and asymmetry parameter, $Q$.}\label{tab:mod}
\end{table*}

\subsection{Dynamical phenomena -- double-periodic pulsation}\label{ssec:dm}
%%%%%%%%%%%%%%%%%%%%%%%%%%%%%%%%%%%%%%%%%%%%%%%%%%%%%%%%%%%%%%%%%%%%%%%%%%%

\begin{figure}
\centering
\resizebox{\hsize}{!}{\includegraphics{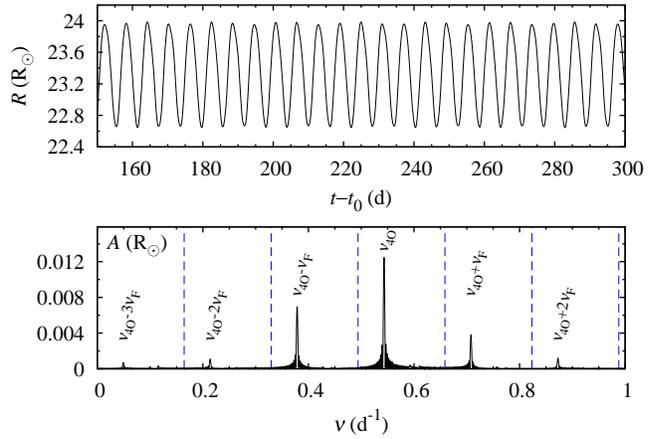}}
\caption{Exemplary model ($M=0.8\MS$, $L=725\LS$, $\Delta T=25$\thinspace K) showing clear beat pulsation; radius variation in the top panel and corresponding frequency spectrum, after prewhitening with the fundamental mode frequency and its harmonics (dashed lines), in the bottom panel. The most significant combination frequencies are labeled.}
\label{fig:beat}
\end{figure}

Six models marked in Fig.~\ref{fig:dynamics} with squares are very interesting. All these models are located $25$\thinspace K on the red side of the blue edge of the fundamental mode instability strip, and clearly show beat pulsation; radius variation for exemplary model is plotted in the top panel of Fig.~\ref{fig:beat}. After fitting the data with sine series including fundamental mode frequency and its harmonics, a strong signal is present in the frequency spectrum of the residual data; its frequency is higher than the fundamental mode frequency by a factor of $\approx\!3.33$ in all the models. After prewhitening with this frequency, we also detect more than hundred of the combination frequencies involving the fundamental mode and the new signal -- see bottom panel of Fig.~\ref{fig:beat}. Investigation of linear period ratios leaves no doubt -- the additional signal corresponds to the fourth radial overtone. In Tab.~\ref{tab:dm} we provide both linear, $P_{\rm 4O}/P_{\rm F}$, period ratios and period ratios from the analysis of non-linear models. Linear and non-linear period of the fundamental mode is also given. We note that except the two independent frequencies and their combinations we find no additional significant signals in the spectra of the beat models (at least down to the amplitude of $10^{-6}\RS$). We have computed additional model sequence of constant luminosity, equal to $700\LS$, with $5$\thinspace K step in the effective temperature, running within $50$\thinspace K from the blue edge of the IS. All models located up to a distance of $25$\thinspace K from the blue edge show the beat pulsation. We conclude that the domain of beat pulsation is adjacent to the blue edge of the fundamental mode instability strip, extends in between $675\LS$ and $800\LS$ and is about $25$\thinspace K wide. 

The fourth radial overtone is a special mode, as pointed out in Sect.~\ref{ssec:linprop}. It is a surface (or strange) mode, trapped in the outer model layers, that can be linearly unstable on the hot side of the fundamental mode instability strip. The red edge of the fourth overtone IS is marked with thick dashed line in Fig.~\ref{fig:dynamics}. For models of set A, the fourth overtone instability strip enters the fundamental mode instability strip; there is a domain in the HR diagram in which both modes are linearly unstable. Here, in principle, the non-resonant beat pulsation can occur, but we have not detected it in our models (neither we specifically searched for it, see below). For models of set B and of $0.6\MS$, the two instability strips just touch, but for models of $0.8\MS$, i.e. for those for which we {\it actually} report the beat pulsation, the two domains are decoupled. The fourth overtone IS approaches the models in which beat pulsation is detected, but for all the models showing the beat pulsation the fourth overtone is linearly weakly damped -- see the linear growth rates reported in Tab.~\ref{tab:dm}. This prompted us to check the consistency between our linear and non-linear code, specially that the linear model properties are very sensitive to the details of numerical mesh or to the values of convective parameters. To this aim we have initialized the non-linear model integration with scaled linear velocity eigenfunction of the fourth overtone, with very small surface amplitude of only $0.01$\thinspace km\thinspace s$^{-1}$ (linear regime). Then, we estimated the growth rate by computing the relative growth of the kinetic energy of pulsation on a cycle-by-cycle basis for the first 40 cycles of pulsation, see \cite{sm08a} for more details. For all six models the growth rates are negative and agree with the linear values to within $\pm 10$\thinspace per cent. We also get excellent agreement for the pulsation periods, and conclude that our linear and non-linear calculations are fully consistent.

\begin{figure}
\centering
\resizebox{\hsize}{!}{\includegraphics{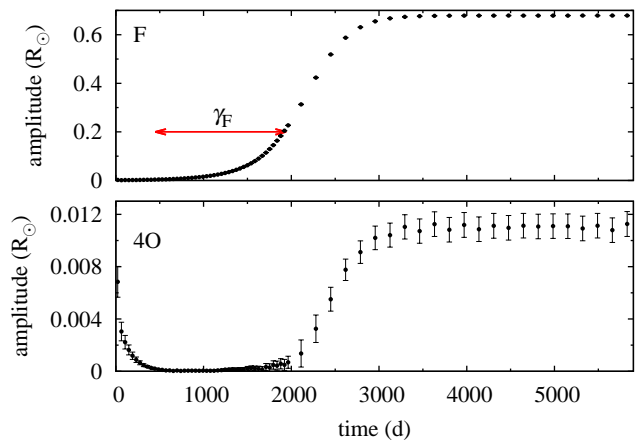}}
\caption{Amplitude of the fundamental mode (top) and of the fourth overtone (bottom) during integration of the model ($0.8\MS$, $675\LS$, $\Delta T=25$\thinspace K) initialized with 4th overtone perturbation.}
\label{fig:dmint}
\end{figure}

How does the beat pulsation arise then? To get more insight into the problem, we carried the model integration starting with pure fourth overtone initial perturbation. Then, using the time-dependent Fourier analysis \citep{kbd87} we extracted the instantaneous amplitudes of the fourth overtone and of the fundamental mode. Results are presented in Fig.~\ref{fig:dmint}. In agreement with linear results, fourth overtone decays at the beginning of the integration. Then, amplitude of the fundamental mode builds-up, first at the rate following from linear analysis (arrow in the top panel of Fig.~\ref{fig:dmint}). The rebirth of the fourth overtone may also be observed then. As soon as amplitude of the fundamental mode becomes significant, i.e. for $t\!>\!2000$\thinspace d, the amplitude of the fourth overtone grows fast, following the growth of the fundamental mode. Finally, the two amplitudes settle at a constant value; the double-mode pulsation is stable. Clearly the fourth overtone {\it lives at the cost} of the fundamental mode. Two mechanisms may transfer the energy between the modes: the resonant and non-resonant mode coupling, which we now investigate. 

The fundamental mode and the fourth overtone are not in the resonance relation. The 3:1 resonance between these modes is close, but the periods do not synchronise at $P_{\rm 4O}/P_{\rm F}=0.333$; the period ratio is $\approx 0.3$ (Tab.~\ref{tab:dm}). Consequently, a mediator is needed for the resonance scenario. We note that parametric resonances were found to play important role in the case of F+1O Cepheid models \citep{sm10} and 1O+2O Cepheid models \citep{ds09}. For the present models we have found only one parametric resonance that is close, $2\nu_{\rm 2O}=\nu_{\rm F}+\nu_{\rm 4O}$. In Tab.~\ref{tab:dm} we provide the mismatch parameter for the models, defined as $\Delta=2\nu_{\rm 2O}-\nu_{\rm F}-\nu_{\rm 4O}$. Loci of the resonance crosses the beat pulsation domain. We note however, that the second overtone is heavily damped in these models (Tab.~\ref{tab:dm}, see also Fig.~\ref{fig:avoidedX}), hence, it is unlikely that the discussed resonance indeed plays a role. 

As discussed in Sect.~\ref{ssec:linprop}, one has to be very careful while using liner period ratios to pinpoint the resonance loci, as non-linear period shifts may be strong. This is not the case for the discussed models however; they are detected directly at the blue edge of the IS and pulsation is weakly non-linear: the growth rates are relatively small, light curves are weakly non-linear (small $R_{21}$; see Fig.~\ref{fig:r21}) and period changes are small (we note that period of the fundamental mode is shorter than the linear value by only $\approx\!0.4$\thinspace per cent, Tab.~\ref{tab:dm}). We see no other candidate for resonant mode interaction that could help to drive the beat pulsation.

We are then left with the non-resonant mechanism behind the beat pulsation, which is counterintuitive, at least when classical Cepheid/RR~Lyrae modeling is recalled. The scenario is commonly discussed in the language of amplitude equations then. For the non-resonant interaction of two modes the long-term behaviour of their amplitudes, $A_1$ and $A_2$, may be described with \citep[see e.g.][]{bg84,sm08b}
\begin{align}
\dot{A}_1&=\big(\gamma_1+q_{11}A_1^2+q_{12}A_2^2\big)A_1,\nonumber\\
\dot{A}_2&=\big(\gamma_2+q_{21}A_1^2+q_{22}A_2^2\big)A_2,
\label{eq:ae}
\end{align}
where $\gamma_i$ are linear growth rates and $q_{ii}/q_{ij}$ are self-/cross-saturation coefficients (all real). The equations were truncated at the cubic terms. The concomitant equations for the phases are irrelevant for the present considerations. The saturation coefficients are complicated functions of stellar structure and mode eigenfunctions \citep[see e.g.][]{vanhoolst}. In practice they may be derived from hydrodynamic trajectories as that displayed in Fig.~\ref{fig:dmint}. It is commonly assumed that the saturation coefficients are negative, and hydrodynamic Cepheid and RR~Lyr models support this assumption \citep[e.g.][]{kbsc02,sm08b}. Consequently, for non-resonant beat pulsation both linear growth rates must be positive, i.e. the two modes must be linearly unstable. This is clearly not the case for the present models. A word of caution is needed. Amplitude equations are valid only if mode growth rates are small as compared to oscillation frequencies (weak non-adiabaticity) and if non-linearities in the pulsation are small. These assumptions are satisfied in classical Cepheid and RR~Lyraes, but likely not in strongly non-adiabatic and non-linear luminous type-II Cepheid models. On the other hand, the discussed models are at the blue edge of the classical instability strip, with small growth rates and relatively low pulsation amplitudes. No strong non-linear effects (significant period changes, period doubling alternations) are detected in these models. The amplitude equations may still be reliable in this regime. The occurrence of beat pulsation requires the positive value of cross-saturation coefficient, $q_{40}$, then, which reflects the driving of the fourth overtone at the cost of the fundamental mode (we have indexed the fundamental mode with `0' and the fourth overtone with `4'). This unusual situation may be a consequence of the special, surface nature of the fourth overtone. Then, assuming that other saturation coefficients are negative it follows from the analysis of amplitude equations that double-mode solution exists and is stable if $\gamma_4-\gamma_0(q_{40}/q_{00})$ is positive. The above expression represents the growth rate of the fourth overtone perturbation in the fundamental mode limit cycle. 

We also note that equations \eqref{eq:ae} may be too simple to describe the pulsation dynamics we have found; inclusion of further, quintic terms may be necessary. Then it might appear that cross-saturation coefficients in \eqref{eq:ae} will remain negative. This will be studied in more detail in a separate paper.

Why does the beat pulsation not occur in other model grids, in particular for models of set A, for which, over a significant fraction of the instability strip, the two modes are linearly excited? First, we cannot claim that beat pulsation is not possible in these domains, as we have not searched for it. This is a tedious and time-consuming task, as for each model several time integrations with different initialization must be carried, see e.g. \cite{sm08b} for more details. 

The discussed form of pulsation was not yet observed. In the models the fourth overtone has small amplitude (Tab.~\ref{tab:dm}), which may hinder its detection. On the other hand, classical pulsators are now frequently observed from space with unprecedented precision (including type-II Cepheids, Plachy et al., in prep). The quality of the dedicated ground-based observations allow to detect mmag signals as well. Therefore, the detection of surface modes may be just a matter of time. The candidates for single-, surface-mode pulsation were already identified among classical Cepheids \citep{ula}. Our models are the first to point that surface modes may be involved in the beat pulsation within classical instability strip. The full understanding of the presented models is thus needed and dedicated study is ongoing.  

\begin{table*}
\begin{tabular}{rrrrrrrrrrr}
         &         & \multicolumn{6}{c}{linear properties} & \multicolumn{3}{c}{non-linear properties} \\
 $L/\LS$ & $\teff$ (K) & $P_{\rm F}$ (d) & $P_{\rm 4O}/P_{\rm F}$ & $\Delta$ & $\gamma_{\rm F}$ & $\gamma_{\rm 2O}$ & $\gamma_{\rm 4O}$ & $P_F$ (d) & $P_{\rm 4O}/P_{\rm F}$ &  $A_{\rm 4O}/A_{\rm F}$ \\
\hline
$675$  & $6222.2$  & $5.650$ & $0.2990$ & $-0.0116$ & $0.033$ & $-1.12$ & $-0.028$ & $5.627$ &  $0.3016$  &  $0.010$ \\
$700$  & $6211.1$  & $5.874$ & $0.2997$ & $-0.0061$ & $0.034$ & $-1.16$ & $-0.013$ &  $5.850$ &  $0.3024$  &  $0.017$ \\
$725$  & $6200.5$  & $6.099$ & $0.3004$ & $-0.0012$ & $0.035$ & $-1.19$ & $-0.007$ &  $6.073$ &  $0.3031$  &  $0.019$ \\
$750$  & $6190.3$  & $6.325$ & $0.3011$ &  $0.0033$ & $0.036$ & $-1.23$ & $-0.008$ &  $6.297$ &  $0.3038$  &  $0.018$ \\
$775$  & $6180.5$  & $6.551$ & $0.3019$ &  $0.0076$ & $0.038$ & $-1.27$ & $-0.015$ &  $6.521$ &  $0.3047$  &  $0.014$ \\
$800$  & $6171.1$  & $6.777$ & $0.3027$ &  $0.0115$ & $0.039$ & $-1.31$ & $-0.025$ &  $6.746$ &  $0.3056$  &  $0.007$ \\
\hline
\end{tabular}
\caption{Linear and non-linear properties of the double-periodic models. $\Delta=2\nu_{\rm 2O}-\nu_{\rm F}-\nu_{\rm 4O}$. All models have $M=0.8\MS$, ${\rm [Fe/H]}=-1.5$ and are located $25$\thinspace K towards the red of the blue edge of the fundamental mode IS.}\label{tab:dm}
\end{table*}

\subsection{Shape of the radius variation curves}\label{ssec:bump}
%%%%%%%%%%%%%%%%%%%%%%%%%%%%%%%%%%%%%%%%%%%%%%%%%%%%%%%%%%%%%%%%%%

\begin{figure}
\centering
\resizebox{\hsize}{!}{\includegraphics{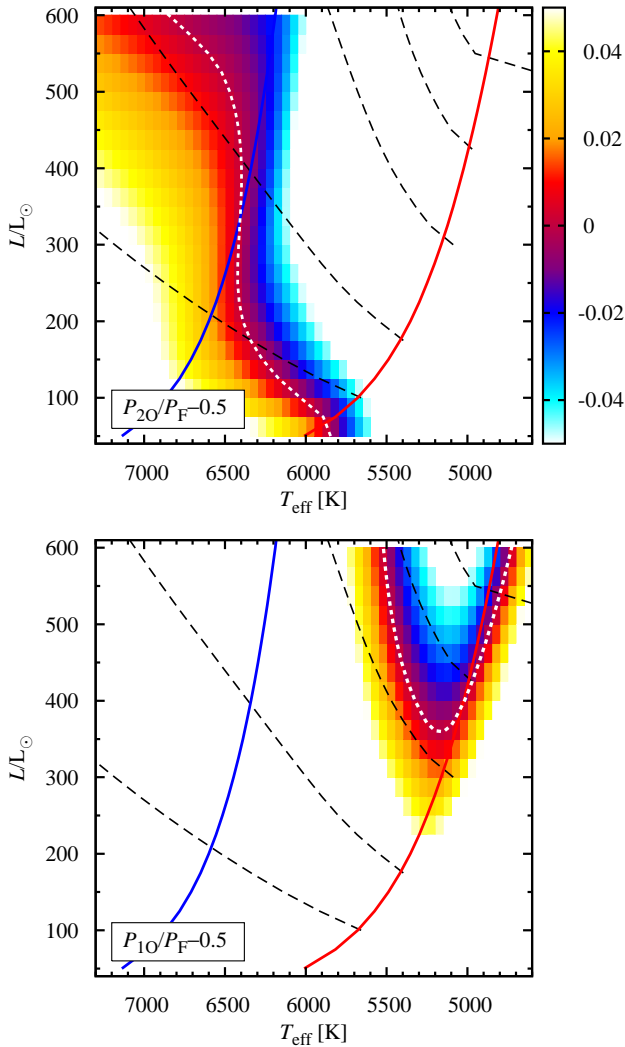}}
\caption{Theoretical HR diagrams with loci of the 2:1 resonances, $\nu_{\rm 2O}=2\nu_{\rm F}$ (top) and $\nu_{\rm 1O}=2\nu_{\rm F}$ (bottom), plotted with dotted lines ($M\!=\!0.6\MS$, ${\rm [Fe/H]}\!=\!-1.5$, set A). Colors code the relevant resonance mismatch parameters defined in the bottom-left part of each panel. Edges of the IS and lines of constant fundamental mode period (dashed lines; from bottom to top 2, 4, 8, 12 and 16\thinspace d) are also plotted. All quantities from linear analysis of static models.}
\label{fig:rez21}
\end{figure}

\begin{figure*}
\centering
\resizebox{\hsize}{!}{\includegraphics{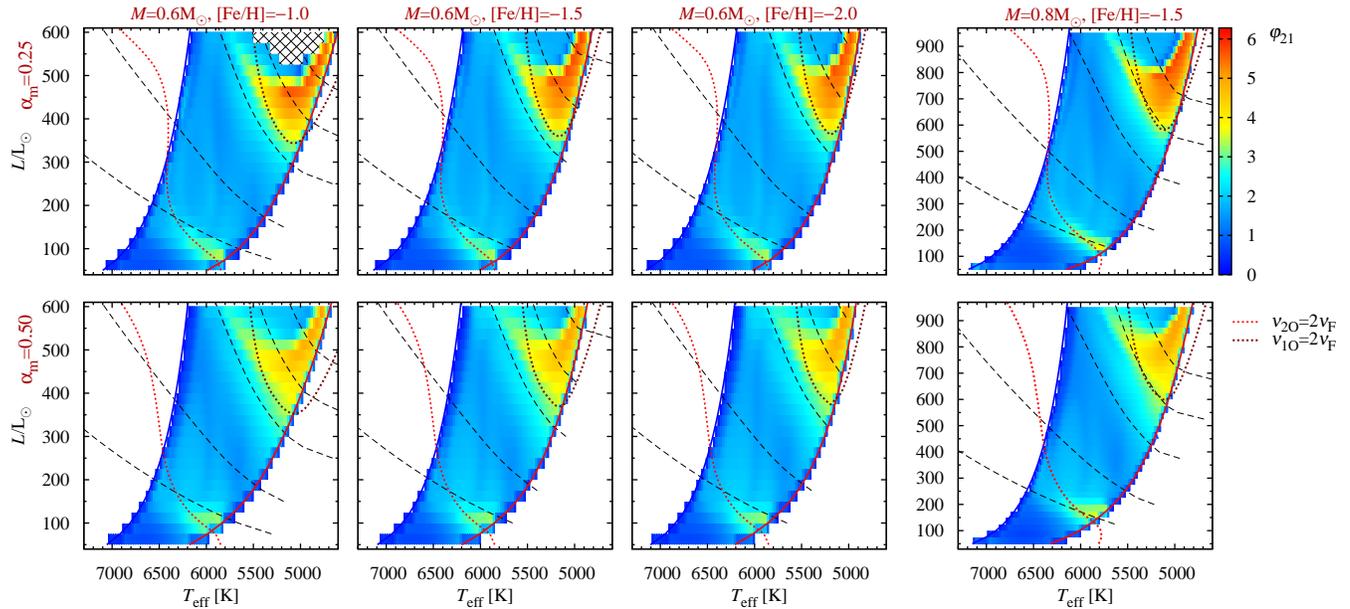}}
\caption{The Fourier phase, $\varphi_{21}$, for the radius variation coded with color as indicated in the top right part of the figure. The loci of the two 2:1 resonances, $\nu_{\rm 2O}=2\nu_{\rm F}$ and $\nu_{\rm 1O}=2\nu_{\rm F}$, are marked with the dotted lines. Dashed lines are lines of constant fundamental mode period equal to (from bottom to top): $2$, $4$, $8$, $12$ and $16$\thinspace d.}
\label{fig:phi21}
\end{figure*}

\begin{figure*}
\centering
\resizebox{\hsize}{!}{\includegraphics{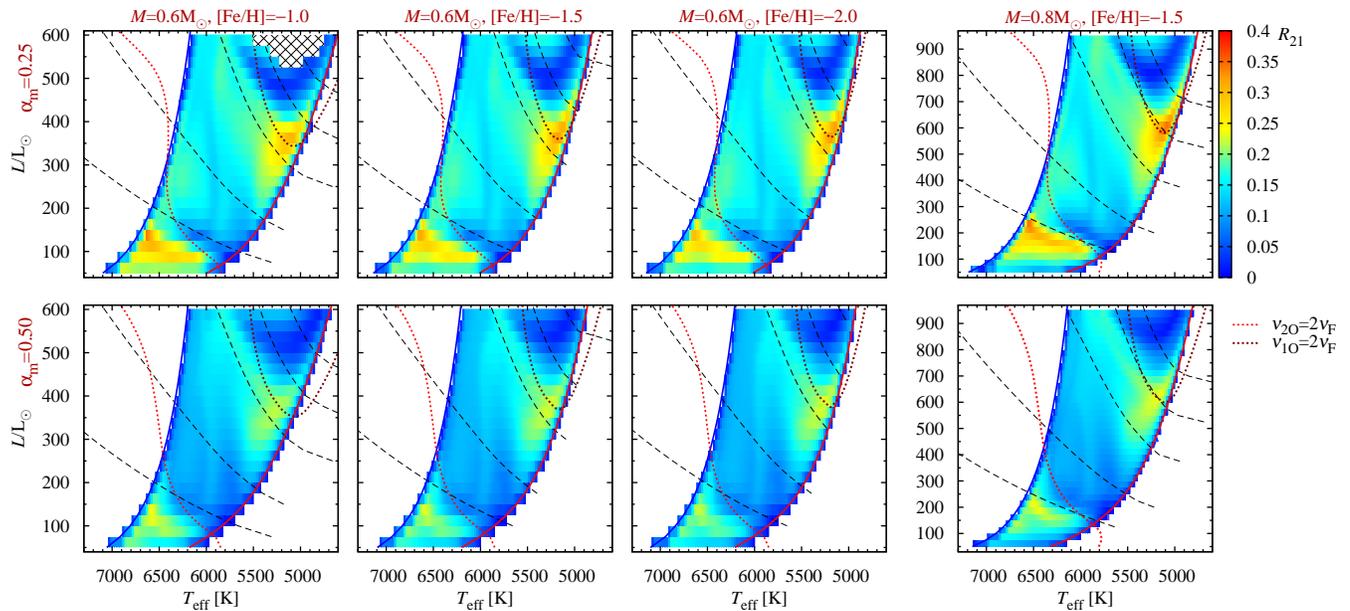}}
\caption{The same as Fig.~\ref{fig:phi21} but for the amplitude ratio, $R_{21}$.}
\label{fig:r21}
\end{figure*}

The 2:1 resonances between the pulsation modes are well known to affect the light-curve shapes of classical pulsators, most notably in the so-called Hertzsprung bump progression. There are two 2:1 resonances that might be in action, between the fundamental mode and the second overtone, $\nu_{\rm 2O}=2\nu_{\rm F}$, and between the fundamental mode and the first overtone, $\nu_{\rm 1O}=2\nu_{\rm F}$. Their loci are plotted in Fig.~\ref{fig:rez21} (for $M\!=\!0.6\MS$, ${\rm [Fe/H]}\!=\!-1.5$, set A models). The first resonance crosses the IS diagonally, at lower luminosities (BL~Her domain). Loci of the second resonance is `V'-shaped, and falls at higher luminosities (W~Vir domain). How far from its center the resonance remains influential is a difficult question that can be answered only through non-linear model integration. The 2:1 resonance between the fundamental mode and the second overtone, that shapes the Cepheid bump progression, and potentially plays a role in so-called binary evolution pulsators \citep{bep_nat}, was most extensively studied. Its resonance band is wide, $0.46\!\lesssim\!P_{\rm 2O}/P_{\rm F}\!\lesssim\!0.55$ \citep{bmk90,bep}. A similar resonance band is found for the 2:1 resonance between the first and fourth overtones that shapes the light curve progression in first overtone Cepheids \citep{aa95,fbk00}. Therefore, in Fig.~\ref{fig:rez21}, a bit arbitrarily we color-code the $P_{\rm 2O}/P_{\rm F}-0.5$ and  $P_{\rm 1O}/P_{\rm F}-0.5$ values in the $(-0.05,\,0.05)$ range to indicate the parts of the HR diagram in which resonances are likely be influential. This Figure should be analysed in conjunction with Figs.~\ref{fig:phi21} and \ref{fig:r21}, in which we plot the maps of the Fourier parameters, $\varphi_{21}$ and $R_{21}$, for the radius variation curves across the IS. While studying Fig.~\ref{fig:phi21} the reader should bear in mind that the Fourier phase, $\varphi_{21}$, is determined with the $2\uppi$ ambiguity. A glimpse at Figs.~\ref{fig:rez21}, \ref{fig:phi21} and \ref{fig:r21} indeed suggests, that the two 2:1 resonances play an important role in shaping the Fourier parameter progressions in type-II Cepheid models. In Sections \ref{sssec.21O2} and \ref{sssec.21O1} the role of the two resonances is discussed in more detail.

The 3:1 resonance between the radial modes is rarely considered in the literature; we are not aware of any observed phenomenon for which the role of the 3:1 resonances was proved crucial. Using the amplitude equation formalism \cite{mb89} studied the general effects of the 3:1 resonance in radial stellar pulsations. They show that this type of resonance affects the pulsation in a very similar fashion to a 2:1 resonance. The progressions of the light/radial velocity curves are qualitatively similar in the case of both resonances; \cite{mb89} concluded that it may be difficult to discriminate between the effect of the two resonances in the observational data. In Section~\ref{sssec.31} we check whether the 3:1 resonances can play a role in type-II Cepheid models.

\subsubsection{The 2:1 resonance with the second overtone}\label{sssec.21O2}
%%%%%%%%%%%%%%%%%%%%%%%%%%%%%%%%%%%%%%%%%%%%%%%%%%%%%%%%%%%%%%%%%%%%%%%%%%%% 
The early radiative calculations pointed that the 2:1 resonance between the fundamental mode and the second overtone may play a crucial role in shaping the light curves of BL~Her-type stars \citep{csv81,kch81}. The extensive analysis was conducted by \cite{bm92} and \cite{mb93} who studied the shapes of the radial velocity, radius and light curves of several sequences of BL~Her models. They found that the 2:1 resonance dominates the dynamical behaviour of their models shaping the progressions of the Fourier decomposition parameters, similar as in the case of classical Cepheids. In the case of the radius variation curve, they found that $\varphi_{21}$ displays a bell-shape progression as period ratio, $P_{\rm 2O}/P_{\rm F}$, is decreased, while $R_{21}$ displays a more complex variation: first it increases, then decreases and gently increases again \citep[see fig.~9 in][]{bm92}. In contrast to Cepheid models however, they found that the strength of the resonance depends sensitively on the stellar parameters, the mass and luminosity. Consequently, the shape of the Fourier progression curves also changes significantly as these parameters are varied.

Our models fully agree with the radiative results outlined above. In Fig.~\ref{fig:phi21} we can easily trace a bell shape in the progression of the Fourier phase, $\varphi_{21}$. The increase of $\varphi_{21}$ clearly follows the $\nu_{\rm 2O}=2\nu_{\rm F}$ resonance loci; the increase is more significant at lower luminosities and weakens as we move along the resonance loci line towards higher luminosities and towards the blue edge. The progression of the amplitude ratio, $R_{21}$, is also clearly correlated with the resonance loci. In Fig.~\ref{fig:r21} we observe a significant decrease of $R_{21}$ as the resonance center is crossed from its hot side towards the cool side, i.e. as $P_{\rm 2O}/P_{\rm F}$ decreases. The effect is less pronounced at higher luminosities, but it is not surprising, as resonance loci approaches the blue edge of the IS, and hence, the pulsation amplitudes, and consequently $R_{21}$, are reduced there.

\begin{figure}
\centering
\resizebox{\hsize}{!}{\includegraphics{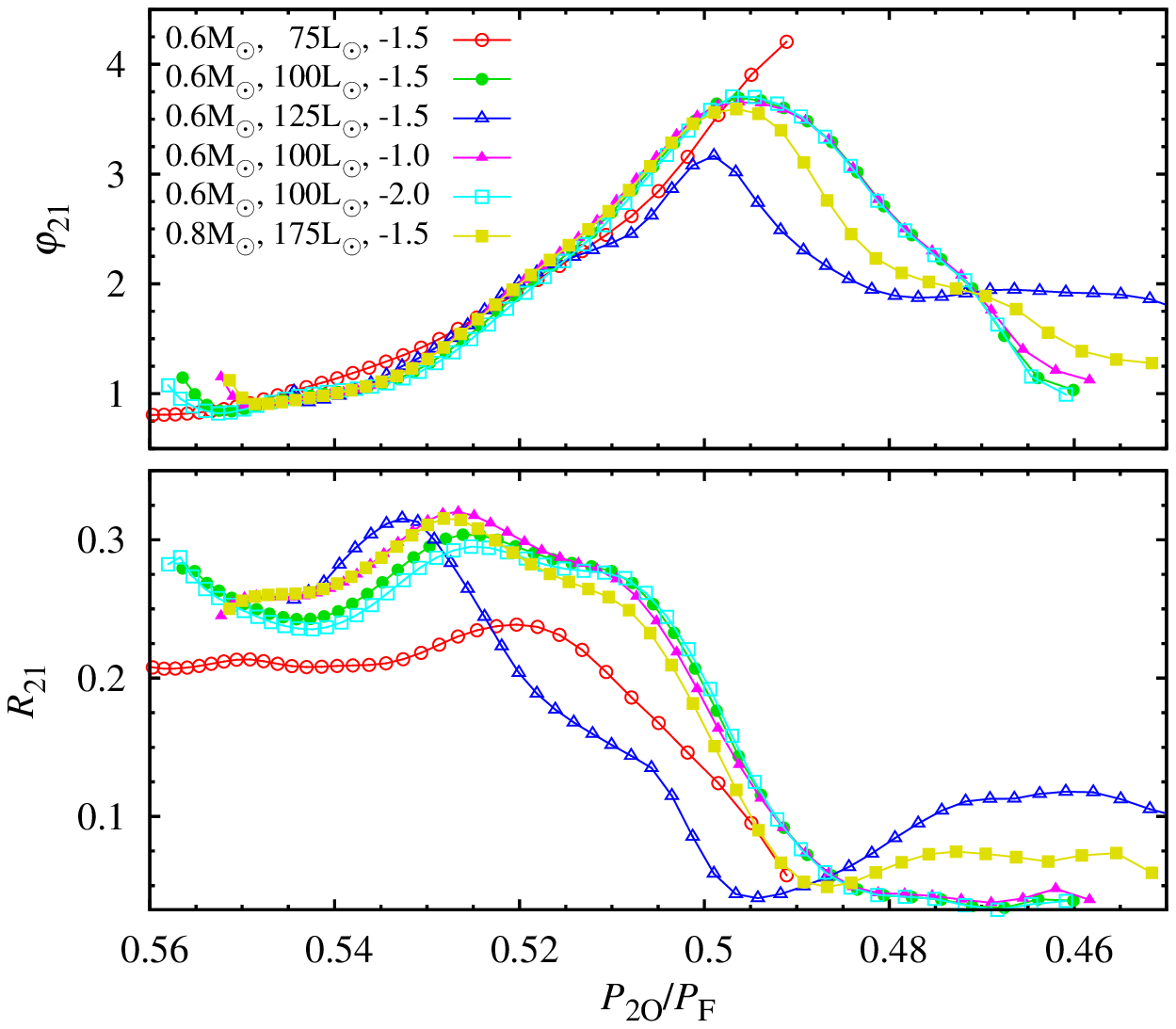}}\\
\resizebox{\hsize}{!}{\includegraphics{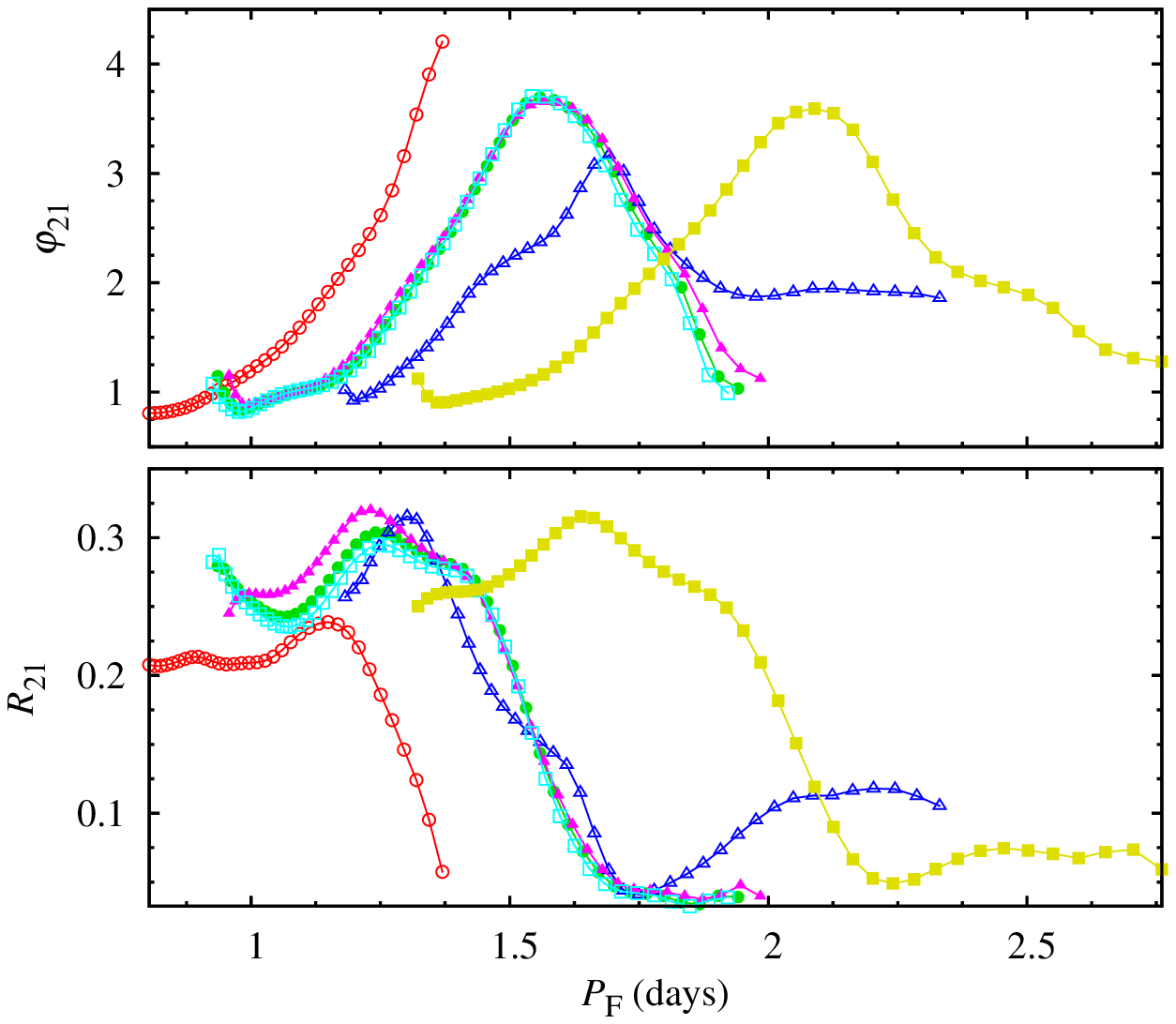}}
\caption{Fourier phase, $\varphi_{21}$, and amplitude ratio, $R_{21}$, for the radius variations versus the $P_{\rm 2O}/P_{\rm F}$ period ratio (top panels) or versus the fundamental mode period, $P_{\rm F}$ (bottom panels). Data for selected model sequences crossing the $\nu_{\rm 2O}=2\nu_{\rm F}$ resonance center are plotted, as indicated in the key.}
\label{fig:fp}
\end{figure}

In Fig.~\ref{fig:fp} we plot the progressions of $\varphi_{21}$ and $R_{21}$ as a function of $P_{\rm F}$ (bottom panels) and of $P_{\rm 2O}/P_{\rm F}$ (top panels) for six horizontal model sequences (of constant $L$). Models of different luminosity, metallicity and mass are plotted, as indicated in the key. The characteristic features, the bell shape in the case of $\varphi_{21}$ progression, and the decrease of $R_{21}$, are shifted with respect to each other, if plotted versus the pulsation period. If period ratio is used on the abscissa instead, the features nicely align, proving the crucial role of the 2:1 resonance in shaping the Fourier parameter progressions.

Our results show that the non-linear shift of the $\nu_{\rm 2O}=2\nu_{\rm F}$ resonance cannot be strong, which is not surprising as non-linear fundamental mode period shifts at lower luminosities are not strong (Sect.~\ref{ssec:pc}, Fig.~\ref{fig:pc}).

Progressions for the light curves are qualitatively different than for the radius variation as analysed by \cite{mb93}. Slow increase of the Fourier phases is expected, rather than bell-shape progression. Such behaviour was found by \cite{t2cep_lmc} who analysed the Fourier parameters for the LMC BL~Her stars.

\subsubsection{The 2:1 resonance with the first overtone}\label{sssec.21O1}
\begin{figure}
\centering
\resizebox{\hsize}{!}{\includegraphics{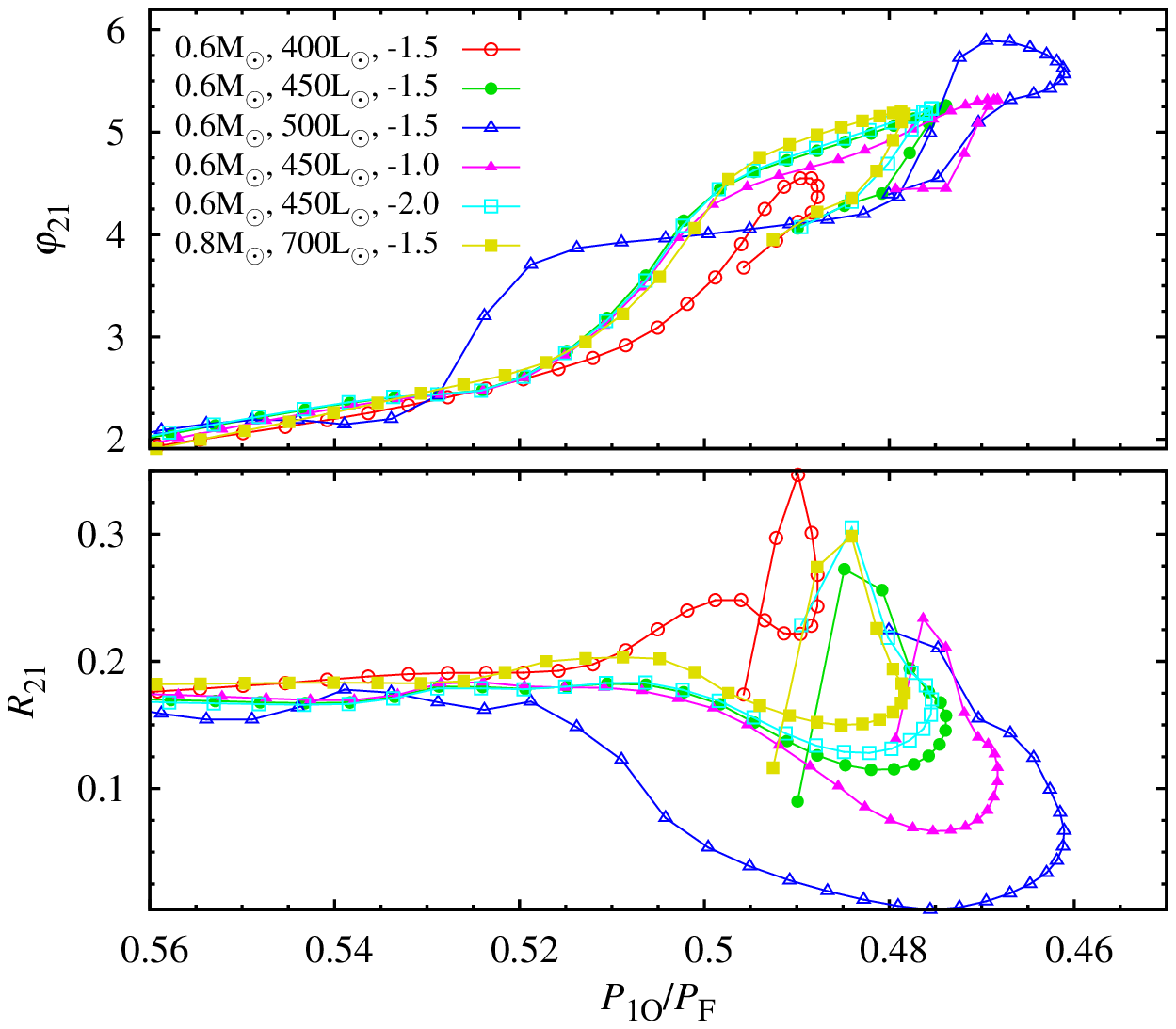}}\\
\resizebox{\hsize}{!}{\includegraphics{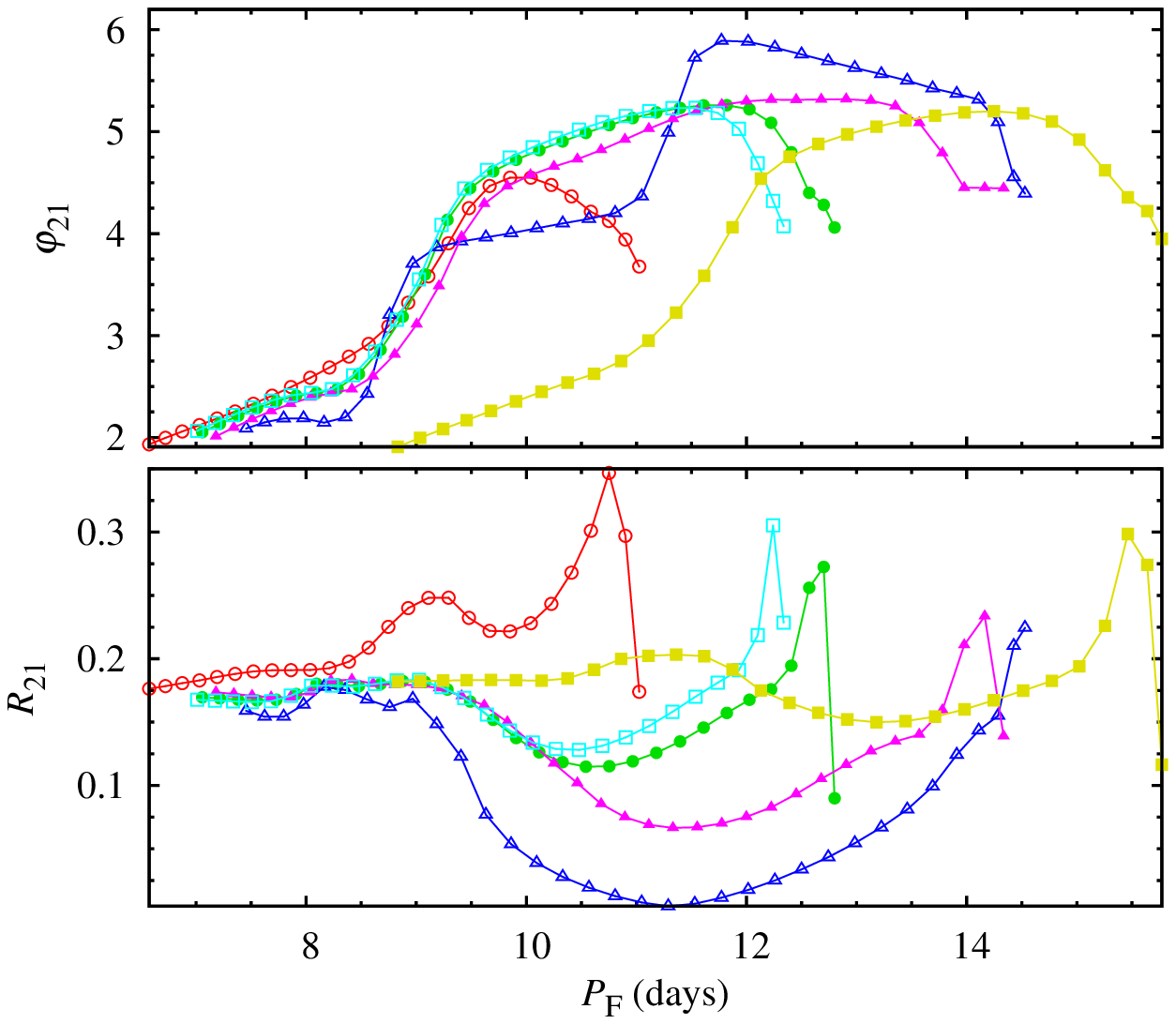}}\\
\caption{The same as Fig.~\ref{fig:fp}, but for model sequences crossing the $\nu_{\rm 1O}=2\nu_{\rm F}$ resonance.}
\label{fig:fp2}
\end{figure}

%%%%%%%%%%%%%%%%%%%%%%%%%%%%%%%%%%%%%%%%%%%%%%%%%%%%%%%%%%%%%%%%%%%%%%%%%%%
The analysis of the potential role of this resonance on the pulsation dynamics is more difficult. Its loci is `V'-shaped (see also Fig.~\ref{fig:rez21}, bottom panel), which means that the period ratio, $P_{\rm 1O}/P_{\rm F}$, does not vary monotonically along a sequence of models with constant $L$. In addition, the cool branch of the resonance is running through, and then on the cool side of the red edge of the IS. Also, at the luminosities considered, strong non-linear period changes (resonance shifts) are expected (Sect.~\ref{ssec:pc}, Fig.~\ref{fig:pc}). At the red edge of the IS we expect that the pulsation amplitude drops, the variation curves become more sinusoidal (less non-linear), which is reflected in the Fourier parameters. Therefore, the effects of the resonance on the models are best visible when models cross the {\it hot branch} of the resonance, which is located well within the IS. Moving along a constant luminosity line towards the red, in Figs.~\ref{fig:phi21} and \ref{fig:r21} we clearly trace the increase of $\varphi_{21}$ and the decrease of $R_{21}$, just as in the case of the 2:1 resonance with the second overtone. Fig.~\ref{fig:fp2} is the analog of Fig.~\ref{fig:fp} for the discussed resonance. Again, the progressions for different model sequences align much better then plotted against period ratio, rather than against pulsation period. The effect is clear but not as pronounced as for the 2:1 resonance with the second overtone; the picture is obscured by the non-monotonic variation of period ratio with the temperature, non-linear period shifts and by the proximity to the red edge. Clearly, the progressions are sensitive to the values of mass and luminosity and to a lesser extent to metallicity.

We note that the potential role of this resonance in shaping the light curves of type-II Cepheids with periods in a range 10--25\thinspace d was noted by \cite{csv81}, based on linear calculations only, and considered cautiously, as no observational hint was available at that time \citep[also, the role of resonances in the Hertzsprung bump progression was disputed, e.g.][]{vs78}. Our non-linear calculations clearly show that this resonance may indeed play a role in shaping the light curves of W~Vir stars at periods as short as 10\thinspace d. In their analysis of type-II Cepheid light curves from the LMC \cite{t2cep_lmc} connect the progressions of $R_{21}$ and $\varphi_{21}$ at periods in between 16 and 20\thinspace d with this resonance (their fig.~8 may suggests that the characteristic progression occurs at even shorter periods, above $\sim$10\thinspace d).

\subsubsection{The effects of the 3:1 resonances}\label{sssec.31}
%%%%%%%%%%%%%%%%%%%%%%%%%%%%%%%%%%%%%%%%%%%%%%%%%%%%%%%%%%%%%%%%%

\begin{figure}
\centering
\resizebox{\hsize}{!}{\includegraphics{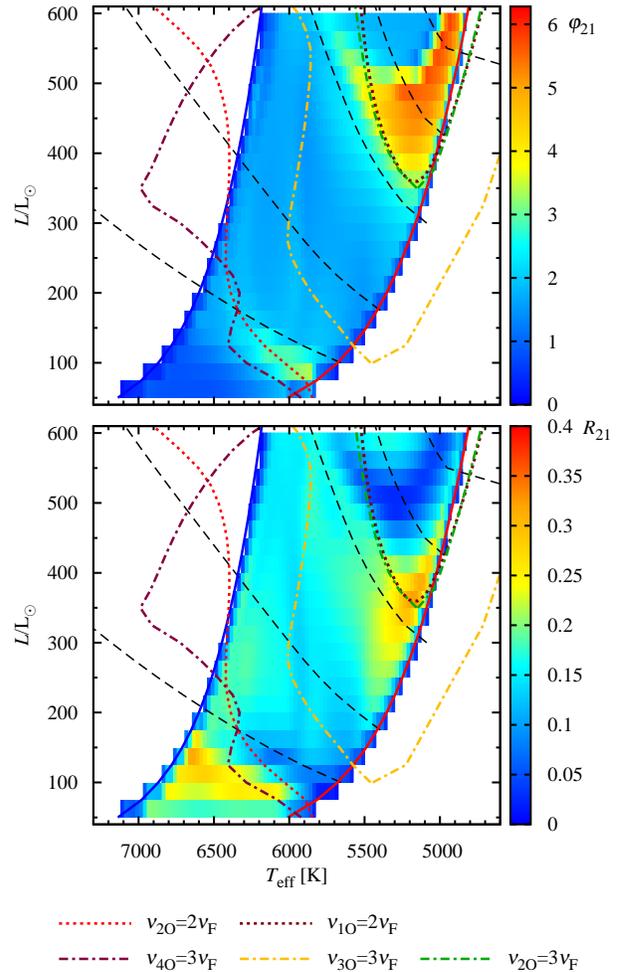}}
\caption{The maps of the Fourier decomposition parameters for radius variation, $\varphi_{21}$ (top) and $R_{21}$ (bottom), across the IS ($0.6\MS$, ${\rm [Fe/H]}=-1.5$, set A). Color scale is the same as in Figs.~\ref{fig:phi21} and \ref{fig:r21}. Overplotted are loci of the 2:1 and 3:1 resonances, as indicated in the key in the bottom part of the Figure.}
\label{fig:rezo31}
\end{figure}

There are three 3:1 resonances that cross the instability strip and in principle may influence the pulsation of our models. To avoid overcrowding in Figs.~\ref{fig:phi21} and \ref{fig:r21}, in Fig.~\ref{fig:rezo31} we overplot the loci of these resonances on top of the Fourier parameter maps ($\varphi_{21}$ in the top panel and $R_{21}$ in the bottom panel) for a model sequence of set A with $0.6\MS$ and ${\rm [Fe/H]}=-1.5$. Results are qualitatively similar for other model sequences considered in this paper. The two resonances, $\nu_{\rm 4O}=3\nu_{\rm F}$ and $\nu_{\rm 3O}=3\nu_{\rm F}$, seem to have no effect on the radius variation curves at all. Situation is more interesting for the $\nu_{\rm 2O}=3\nu_{\rm F}$ resonance. Its loci overlaps with the loci of the 2:1 resonance with the first overtone nearly exactly. The latter resonance was just discussed in the previous Section as responsible for the parameter progressions well marked in the W~Vir domain (see also Fig.~\ref{fig:fp2}). Can $\nu_{\rm 2O}=3\nu_{\rm F}$ also be influential for the high luminosity models? The effects of both 2:1 and 3:1 resonances are qualitatively similar, as shown by \cite{mb89}. In principle, we cannot exclude that the two resonances, 2:1 and 3:1, play a role in shaping the Fourier parameter progressions. The fact that the other two 3:1 resonances play no role in the pulsation indicates however, that the resonant coupling associated with the 3:1 resonances is much weaker than the resonant coupling associated with the 2:1 resonances, at least for the models considered. Therefore we conclude, that the 2:1 resonance, $\nu_{\rm 1O}=2\nu_{\rm F}$, plays a key role in shaping the radius variation curves, while the 3:1 resonance, $\nu_{\rm 2O}=3\nu_{\rm F}$, likely plays no role.

\section{Discussion}
%%%%%%%%%%%%%%%%%%%%%%%%%%%%%%%%%%%%

In our previous studies of BL~Her model sequences, we strongly reduced the eddy viscosity ($\alpha_{\rm m}=0.05$) which led to the large amplitude chaotic behaviour and modulation of pulsation over significant domains in the HR diagram \citep{blher_mod,blher_chaos}. Also, in the largest survey of type-II Cepheid models published so far \citep{kb88} chaotic behaviour was predicted in the low temperature W~Vir models. In general, these results do not agree with the observations. Quasi-periodic modulation of pulsation was not yet observed in type-II Cepheids, while chaotic dynamics, reached through the period doubling cascade, most likely underlies the variability of more luminous, longer period RV~Tau stars. In BL~Her and in W~Vir stars single-periodic pulsation is a rule. Only two stars show the period doubling alternations of low amplitude.

In the present study we kept the eddy viscosity parameter at larger values  ($\alpha_{\rm m}=0.25$ or $\alpha_{\rm m}=0.5$), assuring reasonable pulsation amplitudes. The goal was to check which of the above phenomena persist in more realistic, convective type-II Cepheid models with up-to-date microphysics data. Although in the majority of the models single-periodic fundamental mode pulsation is computed and chaotic dynamics is not present in the models, still more complex dynamical scenarios are possible over significant parts of the instability strip. The most pronounced is the period doubling behaviour, possible in the two large domains within the IS. The low luminosity period doubling domain is the same as first revealed in radiative calculations of \cite{bm92}. The radiative models of \cite {kb88} most likely correspond to the high luminosity period doubling domain. We also detect period-4 pulsation, but no further period doublings and chaos were detected in the models. This is in contrast with \cite{kb88} results for the high luminosity period doubling domain. This is likely due to the fact that as temperature is decreased in the convective model the degree of model's non-linearity diminishes, as we are approaching the red edge of the instability strip. The radiative models simply lack the pulsation quenching mechanism, and are still strongly non-adiabatic and non-linear at low temperatures.

We also note that \cite{dicriscienzo} computed a survey of BL~Her type models but found no period doubling. Their results agree with ours, however. In nearly all their model sequences (see their tab.~1) $L\!<\!130\LS$ and hence, their models run well below the low luminosity period doubling domain. Only in one sequence luminosity is high enough ($L\!\approx\!260\LS$), but the mass is lower ($0.5\MS$) and, since they use different pulsation code than ours, a direct comparison of convective parameters is not possible. 

In very small domain  ($\sim 10$-K wide) within the instability strip we detect small amplitude modulation of pulsation. Other, very interesting phenomenon is double-mode pulsation, in the fundamental mode and in the fourth overtone, which is a surface mode. 

Of the above phenomena, modulation of pulsation and beat pulsation could be missed in the observations, as these forms of pulsation are restricted to narrow domains within the instability strip, and predicted amplitudes of modulation/additional mode are small. On the other hand the models clearly indicate that period doubled pulsation should be common among BL~Her and W~Vir variables, which is not the case. The presence of (so far) single BL~Her-type star and single W~Vir-type star with low amplitude alternations suggests that the computed domains {\it are indeed present}, but clearly their extent and/or amplitude of alternations within must be much lower. Interestingly, the modulation of pulsation is rather easily computed in type-II Cepheid models but not in the less luminous, otherwise similar models of RR~Lyr stars, in which the modulation of pulsation, the Blazhko effect is commonly observed. 

The models we are using are certainly simple. Inclusion of radiation hydrodynamics, finer, dynamical mesh and, most importantly, more realistic 3D treatment of turbulent convection are obvious improvements for the modelling of pulsation in the strongly non-adiabatic, convective and extended envelopes of these luminous stars. The 3D modeling is at its infancy however, only very limited and strongly simplified models of classical Cepheids/RR~Lyr-type stars were computed so far \citep[e.g.][]{geroux,mundprecht}. The significantly better tools are not yet available. 

The amount of good quality observations of type-II Cepheids is growing. The {\it K2} space observations of several type-II Cepheids are expected soon. OGLE-IV is gathering top-quality data for hundreds of these stars \citep{ogleiv}. The phenomena of period doubling, modulation of pulsation or beat pulsation should certainly be searched for. The discovery of the latter phenomenon, beat pulsation, would be very interesting as it would allow to constrain the stellar parameters, just as in the case of beat Cepheids and RR~Lyr stars. In the forthcoming study we plan to investigate the beat pulsation models in more detail and to search the OGLE data for the candidate stars showing this form of pulsation.

\section{Summary}
%%%%%%%%%%%%%%%%%%%%%%%%%%%%%%%%%%%%

We have computed several grids of convective pulsation models of type-II Cepheids. For $0.6\MS$ models three values of metallicity where considered (${\rm [Fe/H]}=-2.0,\,-1.5,\,-1.0$), while for $0.8\MS$ models metallicity was fixed to $-1.5$. Our models cover the BL~Her, W~Vir and only the shortest-period RV~Tau domain. Most important findings are the following:
\begin{itemize}
\item In the most luminous models ($L>600\LS$ for $M=0.6\MS$ and $L\geq 1000\LS$ for $M=0.8\MS$), dynamical instability arises during the model integration, which leads to the decoupling of the outermost model layers and prevents computation of the luminous RV~Tau-type models. This dynamical instability may be connected to pulsation driven mass loss.

\item The models are strongly non-linear. Significant period changes with respect to the linear period values are computed. Both period lengthening and period shortening is possible. For the most luminous models considered, the pulsation period changes may be as large as $\pm 15$\thinspace per cent. These changes however, should not affect the linear period-luminosity relations significantly. 

\item The models predict two large domains in the HR diagram in which period doubling alternations occur.

\item The low luminosity period doubling domain extends vertically in the HR diagram from BL~Her to short-period W~Vir domain. Pulsation periods are in a range $2-6.5$\thinspace d (see Tab.~\ref{tab:pdprop} for more details). It is the same domain as computed by \cite{bm92}, who predicted the existence of period doubled BL~Her stars. Their prediction was confirmed with the discovery of the first period doubled BL~Her star, T2CEP-279 \citep{blher_ogle}. The other star observed by the OGLE project is a strong candidate. In this domain period doubling behaviour is caused by the 3:2 resonance, $2\nu_{\rm 1O}=3\nu_{\rm F}$.

\item The second period doubling domain extends at higher luminosities and longer periods, in the W~Vir domain ($P>9.5$\thinspace d, see Tab.~\ref{tab:pdprop} for more details). Its extent could not be determined, due to dynamic instability preventing computation of very luminous models. The radiative models of \cite {kb88} most likely correspond to the same domain. Period doubling behaviour is likely caused by the 5:2 resonance, $2\nu_{\rm 2O}=5\nu_{\rm F}$. 

\item Double-mode pulsation in the fundamental mode and in the fourth overtone was computed. This form of pulsation was detected only in the more massive ($0.8\MS$) models, adjacent to the blue edge of the IS (fundamental mode periods $\sim5.6-6.8$\thinspace d, period ratio $P_{\rm 4O}/P_{\rm F}\approx 0.3$, Tab.~\ref{tab:dm}). Amplitude of the fourth overtone is small, up to 2\thinspace per cent of the fundamental mode amplitude. The fourth overtone is a surface mode, trapped between the surface and partial hydrogen ionization region. The origin of the beat pulsation is likely non-resonant. Detailed study of these models, search for additional models showing this form of pulsation, as well as analysis of observational data is ongoing and will be reported elsewhere.

\item Single-mode pulsation in the fourth overtone is also possible on the hot side of the instability strip. Expected amplitudes are small however. Properties of such models and prospects of detecting such form of pulsation will be discussed elsewhere. 

\item In very small domain ($\sim 10$-K wide) within the instability strip small-amplitude modulation of pulsation is possible. The models are of W~Vir type, typical pulsation periods are $\sim 9-10$\thinspace d, while modulation periods are several times longer (Tab.~\ref{tab:mod}).

\item The shape of the radius variation curves is affected by the two 2:1 resonances, similar as in the case of Hertzsprung bump progression in classical Cepheids. The resonance with the second overtone, $\nu_{\rm 2O}=2\nu_{\rm F}$, affects the radius variation curves in the BL~Her pulsation domain, while the resonance with the first overtone, $\nu_{\rm 1O}=2\nu_{\rm F}$, is operational in the W~Vir domain.
\end{itemize}

%\clearpage
\section*{Acknowledgements}
This research is supported by the Polish Ministry of Science and Higher Education through Iuventus+ grant (IP2012 036572) awarded to RS. I am grateful to Pawe\l{} Moskalik for reading the manuscript and detailed comments. I also acknowledge fruitful conversations with Wojtek Dziembowski concerning the beat models.

%%%%%%%%%%%%%%%%%%%%%%%%%%%%%%%%%%%%%%%%%%%%%%%%%%

%%%%%%%%%%%%%%%%%%%% REFERENCES %%%%%%%%%%%%%%%%%%

% The best way to enter references is to use BibTeX:

%\bibliographystyle{mnras}
%\bibliography{example} % if your bibtex file is called example.bib

% Alternatively you could enter them by hand, like this:
% This method is tedious and prone to error if you have lots of references

% Don't change these lines
\bsp	% typesetting comment
\label{lastpage}
\end{document}